\documentclass[10.5pt]{article}

\usepackage{mathptmx}
\usepackage{graphicx}
\usepackage{times}
\usepackage{amsmath, amsthm, amssymb}
\usepackage{url}
\usepackage{subfig}
\usepackage{hyperref}
\usepackage{color}

\usepackage{lscape}

\begin{document}

\begin{titlepage}
\begin{center}
\begin{Large}
 Patterns of Emotional Blogging and Emergence of Communities: Agent-Based Model on Bipartite Networks 
\end{Large}

 Marija Mitrovi\'c and Bosiljka Tadi\'c\footnote{e-mail: Bosiljka.Tadic@ijs.si}{}\\

{Department of theoretical physics; Jo\v zef Stefan Institute;
 Box 3000  SI-1001 Ljubljana Slovenia\\ \hspace{1cm}}

\date{March, 2011}
\end{center}

{\bf Abstract:}\\
{\it Background:} We study mechanisms underlying the collective emotional behavior of Bloggers by using the agent-based modeling and the parameters inferred from the related empirical data.

{\it Methodology/Principal Findings:} A bipartite network of emotional agents and  posts  evolves through the addition of agents and their actions on posts.   
The  emotion state of an agent, quantified  by the arousal and the valence, fluctuates in time due to events on the connected posts, and in the moments of agent's action it is transferred to a selected post.  
We claim that the indirect  communication of the emotion in the model rules, combined with the action-delay time and the circadian rhythm extracted from the empirical data, can explain the genesis of  emotional bursts by users on popular Blogs and similar Web portals. The model also identifies the parameters and how they influence the course of the dynamics.

{\it Conclusions:} The collective  behavior is here recognized by the emergence of communities  on the network and the fractal time-series of their emotional comments, powered by the negative emotion (critique). The evolving agents communities leave  characteristic patterns of the activity in the phase space of the arousal--valence variables, where each segment  represents a common emotion  described in psychology.  

\end{titlepage}

\tableofcontents
\newpage

\section{Introduction\label{sec-intro}}
The Internet experience in recent years has revolutionized the  mechanisms that  an individual can exploit to participate in global social dynamics. Consequently, new techno-social phenomena emerge on the Web \cite{kleinberg2008,nature2011,cho2009,castellano2009}, boosting an intensive multidisciplinary research.  In technology research, for example, new generation of services are developing in the direction to integrate  human capabilities in a service-oriented manner \cite{skopik2011}. Behavior of the users in the virtual world has impact on  real-life events, which  becomes a  concern of both social sciences and every day's practice. On the other hand, the data collected from  massive use of the Web  provide the basis to study human behavior  ``experimentally'' at unprecedented scale. For instance, from the high-resolution data stored at various Web portals  (social networks, Blogs, forums, chat-rooms, computer games, etc) information related to user preferences, patterns of behavior, attitudes, and emotions can be inferred for each individual user and user communities gathered around certain popular subjects \cite{vazquez2006,malmgren2009,dodds2009,crane2010,mitrovic2010a,szell2010,gonzales-bailon2010,thelwall2010,mitrovic2010b,mitrovic2011,dodds2011}.

Physics of complex systems and, in particular, the statistical physics of social dynamics, are  focused on the dynamical processes in which human  collective behaviors emerge from large  number of individual actions \cite{castellano2009,vazquez2006,malmgren2009,crane2010}. Combining the concepts of statistical physics with  the machine-learning methods for the emotion detection in texts of messages \cite{thelwall2010,paltoglou2010b}, we have recently performed analysis of large datasets from {\textit{bbcblog.com}} and {\textit{digg.com}} and determined quantitative measures of the collective  behaviors  in which the emotions  are involved \cite{mitrovic2010b,mitrovic2011}. 
 Complementary to our work in Refs.\ \cite{mitrovic2010b,mitrovic2011} where the empirical data are analyzed to extract various complex-systems properties, the present work is a theoretical study  of the processes, underlying the emergence of the collective emotional behavior of Blog users, within the framework of agent-based modeling.

The quantitative analysis of users collective behavior in the empirical data from  \textit{diggs.com} and \textit{bbcblog.com} in Refs.\ \cite{mitrovic2010b,mitrovic2011} has been enabled by mapping the high-resolution data onto bipartite networks of users and posts, as two natural partitions. The idea of bipartite networks makes the ``firm ground'' also in the present theoretical model, where the agents interact indirectly over the posts.  We also make use of 
several other features, observed in various empirical data, that  are relevant for  designing the dynamic rules of the theoretical model: 
\begin{itemize}
\item {\it Universality  of user's behavior} related with the action-delay and the circadian cycles \cite{crane2010,vazquez2006,mitrovic2010a,malmgren2009};
\item {\it User communities} occurring in the cyberspace are  reminiscent to the ones in real life, however, different time scales and grouping mechanisms might be involved \cite{mitrovic2010a,mitrovic2010b,lancichinetti2010,mitrovic2011};
\item {\it Quantitative measures of emotions} have been introduced in psychology research \cite{scherer2005}. In particular, based on Russell's multidimensional model of affect \cite{russell1980}, each known emotion can be represented by a set of numerical values in the corresponding multidimensional space.   
Two fundamental components of emotion, to which we  refer in this work,  are the arousal, related to reactivity to a stimulation,  and the valence, measuring intrinsic attractiveness or aversiveness to a stimulation. These components of 
emotion can be measured  in laboratory based on the related psychophysiological and neurological activity \cite{emotions-Oxford2007,calvo-review2010}. Moreover, a systematic association has been recognized \cite{yarkoni2010} between individual emotional characteristics and word use. The arousal and valence components of an emotion can be retrieved from a written text by suitable machine-learning methods,  which are being developed for a specific type of data  \cite{calvo-review2010,nikolauIEEE2011,paltoglouIEEE2011}. 
\end{itemize}
 Systematic analysis of the patterns of user behaviors and the emotion contents in the texts of comments in the empirical data from popular Blogs \cite{mitrovic2010b},  discussion-driven Diggs \cite{mitrovic2011}, and Forums \cite{warsaw2011}, suggests that negative emotions (critique) drive the  activity on these Web portals. 
However, the   mechanisms working behind this global picture have not been well understood. 
In order to elucidate the role of emotions in the blogging interactions, and to point out potential parameters and levels where the process can be controlled, we devise an agent-based model. The agents are spreading their emotions in a bipartite network environment. The agent's properties, the rules and the parameters of the model are closely related with the empirical data from Blogs and Diggs. 

Agent-based modeling \cite{schweitzer2007,gilbert2008,castellano2009}, where different properties of agents influence their actions, provides suitable theoretical framework   for numerical simulations of social phenomena. 
Recently a model for product-review with the emotional agents in a mean-field environment has been introduced \cite{schweitzer2010}, with the agents emotional states described by two state variables $(a_i,v_i)$. These variables correspond to the psychological values of the arousal and the valence, respectively, in view of the Russell's two dimensional circumples model  \cite{russell1980,barrett1998,scherer2005}.

Building on the Ref.\ \cite{schweitzer2010}, here we study an agent-based-model to explore the emergence of user communities on Blogs, where the emotional contents are communicated {\it indirectly  via comments} that they leave on the posts. For this purpose, our emotional agents are situated on a {\it weighted bipartite network} consisting of users (agents) and posts, where by definition no direct link between the nodes of the same partition is allowed. The weighted links between the nodes of different partitions represent the number of comments of an agent to the linked post. 
Motivated by the realistic situation on  Blogs, in our model the network itself evolves over time due to the arrival of new users and the addition of new posts, and due to user's actions on previous posts of their preference.
The emotional state, measured by the arousal and the valence variables,  which are attached to each agent-node, 
is influenced by time-varying fields on the posts surrounding that agent on the  evolving network.   As in real-life, the elevated arousal may induce an action of the emotional agent on a post, according to the rules introduced  below in section\ \ref{sec-model}. In the moment of  action on a post, the agent's current emotion  arousal and valence components are transferred to the comment that the agent leaves on that post,  where it can be experienced  by other agents.  
Thus the fields themselves evolve with the network evolution and differ for each agent, depending on its position on the network. 
In order to have realistic dynamics, we design the rules of actions that are motivated by systematic observations of the activity patterns in  the   empirical data at Blogs and Diggs, as described in section \ref{sec-data} and in the supportive information. Moreover, the set of  parameters that control the dynamics of our model are inferred from the empirical  data  of popular discussion-driven Diggs, as explained below.

\section{Materials and Methods}
\subsection{Datasets\label{sec-data}}
In this work we use  large datasets\footnote{Data accessible on http://www.cyberemotions.eu/data.html under terms and conditions of the CYBEREMOTIONS project.} related to popular posts, from which  we (i) study  the temporal  patterns of events, that motivate the dynamic rules of our model, and  (ii) extract  realistic values of certain control parameters of the model.
 As it will be clear below, the present analysis aims for different features  of these empirical datasets, yielding several new results in comparison with the ones presented in our previous study \cite{mitrovic2010b,mitrovic2011}.

 The datasets that we use are collected from \textit{bbcblog.com} and \textit{diggs.com} \cite{mitrovic2010b,mitrovic2011} and have  high temporal resolution, information about identity of each user and of each post, as a unique ID, and precise relationship between users and comments-on-posts, as well as  full text of all posts and comments. The subsets of data related with {\it popular posts}  as posts with more than 100 comments are selected together with all users linked to them, as good candidates for the analysis of collective behavior of users \cite{mitrovic2010b,mitrovic2011}. In addition, \textit{diggs.com} data contain information about comment-on-comment.  Thus, we select the subset of popular posts,  termed discussion-driven Diggs (ddDiggs),  on which more than 50\% of comments represent reply to the comments  of other users. This data consists of  $N_P=3984$ discussion-driven Digg stories, on which $N_C=917708$ comments are written by $N_U=82201$ users \cite{mitrovic2011}.  
In addition, texts of posts and comments are classified by machine-learning methods with the {\it emotion classifier}  designed in Ref.\ \cite{thelwall2010} and trained at Blog-type of texts. 
 Using the emotion classifier the texts are designated  as carrying either positive or negative emotion valence, or otherwise are neutral \cite{mitrovic2010b}.

From the discussion-driven Diggs dataset here we analyze the temporal patterns of activity related to both users and  posts. Parts of these patterns are shown in Figs.\ \ref{fig-ddDiggs_patterns}a,b.  Each user (post) occurring in that dataset is given a unique index, plotted along vertical axis, sorted by the time of  first appearance in the dataset.  For each user index,  points along time axis indicate the times when an activity of that user  occurred to anyone of the posts. Analogously, the points on the posts pattern indicate the times when an activity occurred at that post by anyone of the users. 
In the post-activity pattern, shown in Fig.\ \ref{fig-ddDiggs_patterns}b, 
dense points in a narrow time window following the post appearance time indicate an intensive activity at that post. This might be related with certain {\it exposure of the posts} to users during that time period. The width of the exposure time window, $T_0$, will be recognized as a relevant parameter in the dynamics.
Whereas,  different type of the dynamics  beyond the exposure window is manifested in systematically reduced activity until eventually the post ceases to be active (expires).  

The situation is entirely different  when looked from the point of view of the users. The patterns of activity of every user over time is shown  in Fig.\ \ref{fig-ddDiggs_patterns}a. The user indexes are ordered by the time of their first appearance in the dataset, hence   the top boundary of the plot
indicates the appearance of new users, relative to the beginning of the dataset. The profile of the top boundary shows that new users arrive in ``waves''. Moreover, the arrivals of new users boost the activity of previous users, which is manifested in the increased density of points in depth of the plot below each ``wave''.  This feature of the dynamics is utilized for designing  the model rules in section\ \ref{sec-model}. Further quantitative measures of the temporal patterns in Figs.\ \ref{fig-ddDiggs_patterns}a,b are given in relation to the model parameters in section\ \ref{sec-parameters}.

The {\it number of newly arrived users}  (with respect to the beginning of the dataset) in  suitably defined time bin can be readily extracted from the dataset. The time series of new users $p(t)$ per time bin (in units of $t_{bin}=5$ minutes), inferred from the ddDiggs dataset  is shown by the red line in Figure\ \ref{fig-ddDiggs_userts}.
It exhibits characteristic {\it daily cycles} superimposed on the fractal fluctuations with long-range correlations.  Note that these cycles are related with the occurrence of ``waves'' in the activity patterns in Fig\ \ref{fig-ddDiggs_patterns}a. The signal has the  power-spectrum  of the type $S(\nu) \sim 1/\nu^\phi$,  with $\phi\approx 1.5$, shown in top panel in  Fig.\ \ref{fig-ddDiggs_userts}.  The time series of the {\it number of all active users} per time bin, extracted from the same dataset is shown in the same Figure \ref{fig-ddDiggs_userts} by the green line. It has a similar fractal structure and the power-spectrum of $1/\nu$-type. The power spectrum is correlated over the range of frequencies,   which correspond  to times larger than 2 hours in the time domain. Further analysis of this dataset, which is relevant for this work, is given  in respect to extracting the control parameters of the dynamics in section\ \ref{sec-parameters}.
Detailed analysis of the emotion contents of the comments  and the related time series of the emotional comments can be found in Ref.\ \cite{mitrovic2011}. In the supporting information, Figure\ S1 shows the time-series of the number of emotional comments with positive (negative) valence from the same dataset which is analyzed here, and the excess of the negative emotions.

\subsection{Dynamics: Model of Emotional Bloggers on Evolving Bipartite Network\label{sec-model}}
In the spirit of agent-based modeling, the agents (representing users on Blogs) are given certain properties that may affect their actions, i.e., the dynamic rules of the model. Conversely, these agent's properties are changed due to dynamic interactions between them, which imply further changed actions, and so on. To describe emotional actions of users on Blogs, we adapt the agents whose emotion states are described by two variables---arousal and valence, first introduced in Ref.\ \cite{schweitzer2010}. Apart from the emotion variables, the agents in our model have additional properties indicated below in Eq.\ (\ref{eq-2agents-def}), and are subjected to the dynamically active networked environment, which affects their actions. Moreover, thy are designed for multiple actions in course of the dynamics, thus contributing to the emergence of collective emotional behavior. 

As mentioned above, here the emotional agents are adapted to interact {\it indirectly} via posts on a bipartite network of the agents and posts. Thus the essential elements of our model are: 
\begin{itemize}
\item {\it 2-dimensional local maps}, describing the emotion variables arousal and valence $\{a_i(t),v_i(t)\}$ of each agent; 
\item {\it Interaction environment}, represented by an {\it evolving bipartite network} of agents and posts, through which the agents emotion is spreading;
\item {\it Driving noise}, applied to  systematically perturb the system boosting its internal dynamics. In our model the system is driven by adding new users, according to the time-series  $p(t)$ of new users, which is inferred from the empirical data of ddDiggs, as explained above. No other  type of driving is considered in this work.
\end{itemize} 

It should be stressed that the bipartite network type, with the users (agents) and the posts as two partitions and the weighted links between them representing the number of comments of the user to the post, is necessary in order to take into account the fundamental nature of the dynamics on Blogs: users, as the nodes of the same partition, are never connected directly, but only through the posts. Technically, both types of nodes appear as the {\it objects} at the same level. Therefore, we actually have two types of ``agents'', user and post agent, with different properties indicated here:
\begin{equation}
U[g; v_i(t),a_i(t); Lists_i...,\Delta t] \ ;~~~~~~~~~~~ P[t_P; <v_p(t)>, 
<a_p(t)>; Lists_p...] \ .
\label{eq-2agents-def}
\end{equation}
The dynamical variables arousal and valence $a_i(t)$ and $v_i(t)$ are properties of the user nodes,  which vary in time as explained in detail below. In analogy to  real Blogs, at the moments of actions the emotion (arousal and valence) of the user are transferred to the post (precisely to the comment that the user puts on the post), and thus contribute to the current overall emotional content on that post, $<v_p(t)>,<a_p(t)>$. Both user and post objects have {\it individual Lists} of connections on the evolving bipartite network, which are updated through the user actions, as explained below in section \ref{sec-rules}. Additional properties that may affect the dynamics in our model  are the post life-time, $t_P$, and the user action-delay, $\Delta t$, as well as the user probability for posting a new post, $g$. These properties can be inferred from the considered dataset, as done below in section \ref{sec-parameters}.
 
Beside the dynamic rules, which are motivated by the blogging processes, other gross features of our model are different from previous work on the emotional agents in Ref.\ \cite{schweitzer2010}. The relevant differences are due to the role of networked environment and the endogenous driving, without any external noise.  
Moreover, at the level of individual agents, apart form the two emotion variables which are given by the same type of the nonlinear maps Eqs.\ (\ref{eq-arousal}-\ref{eq-valence}) as in Ref.\ \cite{schweitzer2010}, the agents in our model have additional properties which  affect their actions. These are the inclination towards posting new posts and the action delay, measured by the quantities $g$ and $\Delta t$, respectively, as well as the lists of connections (posts) on the network, which are unique for each agent. Hence the environmental fields in  Eqs.\ (\ref{eq-arousal}-\ref{eq-valence}) are practically different for each agent on the network.
In this way the network environment  induces  (and keeps track of) the heterogeneity among  the agents in a natural way,  through  the lists of posts to which they were connected in the course of their actions. Having these general remarks, we introduce details of the model in the remaining part of this section.
 We first explain the dynamic rules of the model and define all the parameters that control the dynamics.

\subsubsection{Emotional states of individual agents\label{sec-individual}}
Following Ref.\ \cite{schweitzer2010}, we assume that the individual emotional state  (arousal and  valence) of each agent can be described by two nonlinear equations, which are subject of the environmental fields. For our system on bipartite networks, the arousal and the valence are associated with each user-node(!) and their values, kept in the intervals $a_i(t)\in [0,1]$ and $v_i(t)\in [-1,1]$, are updated according to the following nonlinear maps:
\begin{equation}
a_{i}(t+1) = \left\{ \begin{array}{ll}
 (1-\gamma_{a})a_{i}(t) +[h^{a}_{i}(t)+qh^{a}_{mf}(t)](d_{1}+d_{2}(a_{i}(t)-a_{i}(t)^
{2}))(1-a_{i}(t))    &\mbox{if $\Delta t_{i}<1$}\\
(1-\gamma_{a})a_{i}(t)
&\mbox{otherwise} 
       \end{array} \right.
\label{eq-arousal}
\end{equation}
and
\begin{equation}
v_{i}(t+1) = \left\{ \begin{array}{ll}
 (1-\gamma_{v})v_{i}(t) +[h^{v}_{i}(t)+qh^{v}_{mf}(t)](t)(c_{1}+c_{2}(v_{i}(t)-v_{i}(t)^
{3}))(1-|v_{i}|)           &\mbox{if $\Delta t_{i}<1 $}\\
(1-\gamma_{v})v_{i}(t)
&\mbox{otherwise} 
       \end{array} \right.
\label{eq-valence}
\end{equation}
where $i=1,2\cdots N_U(t)$ indicates the index of user node and $t$---the time bin. 
The coefficients $d_1,d_2$ and $c_1, c_2$ characterize the maps themselves, while the network environment effects appear through two types of fileds: the local fields $h^{a}_{i}(t)$ and $h^{v}_{i}(t)$, and the mean fields $ h^{a}_{mf}(t)$ and $ h^{v}_{mf}(t)$. Note that the local fields $h^{a}_{i}(t)$ and $h^{v}_{i}(t)$ vary not only in time but also  from user to user, depending on their connections on the network, and due to evolution of the network itself (see details below). Whereas, the mean fields $ h^{a}_{mf}(t)$ and $ h^{v}_{mf}(t)$ may act on a larger number of users, while also fluctuating in time. In our model they steam from a currently active posts and, thus, may be seen by all users who are attached to these posts. The mean fields indicate how the overall activity moves through  posts, as a kind of   ``atmosphere'' at the Blogsite. The contribution from the mean fields in our model is taken with a fraction $0 \leq q\leq 1$, which is varied as a free parameter in Eqs.\ (\ref{eq-arousal}) and (\ref{eq-valence}), and it is added to the contributions from the local fields.  
A user-node receives the stochastic inputs from  the network  in certain instants of time, when the events occur in its network surrounding, and reacts to them with a delay.   The {\it delay time} $\Delta t$ of each user is counted continuously. When the delay time is smaller than the computational time bin $\Delta t < 1 ~t_{bin}$, the user is {\it prompted for the update} of its arousal and valence according to the full expressions indicated by top lines in the Eqs.\ (\ref{eq-arousal}) and (\ref{eq-valence}). Otherwise, the arousal and valence values are only relaxing, with the rates $\gamma^a=\gamma^v\equiv \gamma$. 

For  understanding the dynamics at local level, let us consider for a moment  the Eqs.\ (\ref{eq-arousal}) and (\ref{eq-valence}) as  two-dimensional nonlinear maps of an {\it isolated node} subjected to a given constant field. 
In Fig.\ \ref{fig-maps} we show the situation for several values of the field pre-factors, while  keeping all other parameters fixed to the values which are used later in the simulations. Depending on the parameters, the maps can reach different fixed points. In particular, the arousal map always leads to an attractive fixed point, the position of which depends on the strength of the field---larger arousal is reached when the field is stronger, cf. Fig.\ \ref{fig-maps}a.
In the case of valence, two fixed points can be reached, one at the positive valence, when the fields are positive (upper branch),  and the other one in the area of negative valence, which is attractive when the fields are negative (lower branches in the Figure\ \ref{fig-maps}b). 
In general, when the nonlinear maps are coupled on a network, the network environment affects each individual map through a feedback loop, causing synchronization  \cite{arenas2008,buzna2009} or other self-organization effects \cite{levnajic2009} among the nodes. In our case the network affects 
the fields,  which thus fluctuate at every time step and depending on the node's particular position on that network. The dynamics of the fields can thus be visualized in the local map as jumping of the trajectories $v_i(t),a_i(t)$ from one  branch of the map to another branch, and consequently, being attracted to another area of the phase space.  As will be explained in detail below in section \ref{sec-rules}, the nonlinear mapping takes part only when the agent is prompted to act. Meanwhile, the maps are only relaxing towards the origin.

\subsubsection{Agents interaction and network evolution: Rules \& Implementation\label{sec-rules}}

The fields $h^{a}_{i}(t)$ and $ h^{a}_{mf}(t)$ in Eq.\ (\ref{eq-arousal}), which affect user $i$ arousal at step $t+1$, are determined from the posts in the {\it currently active part of the network}, ${\cal{C}}(t,t-1)$, along the links of that user. Specifically, 
\begin{equation}
 h^{a}_{i}(t)=\frac{\sum_{p\in{\cal{C}}(t,t-1)}A_{ip}a_{p}^{\cal{C}}(t)(1+v_{i}(t)v_{p}^{\cal{C}}(t))
}{\sum_{p\in{\cal{C}}(t,t-1)}A_{ip}n_{p}^{\cal{C}}(t)(1+v_{i}(t)v_{p}^{\cal{C}}(t))}; \quad
h^{a}_{mf}(t)=\frac{\sum_{p\in{\cal{C}}(t,t-1)}a_{p}^{\cal{C}}(t)}{\sum_{p\in{\cal{C}}(t,t-1)}n_{p}^{\cal{C}}(t)
} \ ,
\label{eq-ha-fields}
\end{equation}
where  $a_{p}^{\cal{C}}(t)$ and $v_{p}^{\cal{C}}(t)$ are the total arousal and the average valence of the
post $p$ calculated from the comments in two preceding time steps, while $n_{p}^{\cal{C}}(t)$ is the number of all 
comments  posted on it during that time period. $A_{ip}$ represents the matrix elements of the network,
i.e., $A_{ip}>0$ if user $i$ is  connected with the active post $p$, while $A_{ip}=0$ if there is no link between them at the time when the fields are computed. Note that such links may appear later as the system evolves!
In Eq.\ (\ref{eq-ha-fields}) the individual arousal fields $h^a_i(t)$ is  modified by (dis)similarity in user's actual valence, $v_i(t)$,  and the valence of recent comments on the post, $v_{p}^{\cal{C}}(t)$.

Regarding the valence fields in Eq.\ (\ref{eq-valence}), we take into account 
contributions from the positive and the negative comments separately, while the neutral comments  do not contribute to valence field. Depending on the current emotional state of the 
agent,  positive and negative fields can lead to different effects
\cite{schweitzer2010}, in particular, positive (negative) state will be influenced more with negative (positive) field, and vice versa.
Here we assume that both components influence user valence, 
but with different strength according to the following expression:
\begin{equation}
 h^{v}_{i}(t)=\frac{1-0.4r_{i}(t)}{1.4}\frac{\sum_{p\in{\cal{C}}(t,t-1)}A_{ip}N^{+}_
{p}(t)}{\sum_{p\in{\cal{C}}(t,t-1)}A_{ip}N^{emo}_{p}(t)}-\frac{1+0.4r_{i}(t)}{1.4}
\frac{\sum_{p\in{\cal{C}}(t,t-1)}A_{ip}N^{-}_{p}(t)}{\sum_{p\in{\cal{C}}(t,t-1)}A_{ip}N^
{emo}_{p}(t)} \ ,
\label{eq-hv-fields}
\end{equation}
where  the valence polarity of the user $i$ is given by  
$r_{i}(t)=\frac{v_{i}(t)}{|v_{i}(t)|}$, and 
$N^{\pm}_{p}(t)$ is the number of positive/negative comments written on
post $p$ in the period $[t-1,t]$. The normalization factor $N^{emo}_{p}(t)$ is
defined as 
$ N^{emo}_{p}(t)=N^{+}_{p}(t)+N^{-}_{p}(t)$. 
The mean-field contributions to the valence steam from the entire set of currently active posts ${\cal{C}}(t,t-1)$, and are independent on how users are linked to them:
\begin{equation}
 h^{v}_{i,mf}(t)=\frac{1-0.4r_{i}(t)}{1.4}\frac{\sum_{p\in{\cal{C}}(t,t-1)}N^{+}_{p}(t)
}{
\sum_{p\in{\cal{C}}(t,t-1)}N^{emo}_{p}(t)}-\frac{1+0.4r_{i}(t)}{1.4}\frac{\sum_{
p\in{\cal{C}}(t,t-1)}N^{-}_{p}(t)}{\sum_{p\in{\cal{C}}(t,t-1)}N^{emo}_{p}(t)} \ .
\label{eq-hvmf}
\end{equation}
 However, the mean-field effects  are  perceived individually by each user, depending on the polarity $r_i(t)$ of user's current valence. 

{\bf The rules of agents interactions on the network} are formulated in view of user behavior on real Blogs and Diggs and  the observations from the quantitative analysis of the related empirical data. In particular,  the dynamic rules are motivated by the temporal patterns in Figs.\ \ref{fig-ddDiggs_patterns}a,b and the time-series in Fig.\ \ref{fig-ddDiggs_userts}, indicating how the number of active users arises in response to the arrival of new users. Moreover,  additional features of the dynamics on ddDiggs, shown in supporting information, Figure\ S1 and Movie\ S2, suggest the dynamics with the dominance of negative emotions and with user's focus systematically shifting towards different posts. 
 In the implementation of the model rules, we also make use of some   general features of human dynamics, i.e., the occurrence of circadian cycles and delayed action to the events, mentioned in the Introduction, and  assume that the arousal drives an action, as commonly accepted in the psychological literature.

The rules are implemented in the C++ code as follows. 
 The system is initialized with typically 10 Users who are connected to 10 Posts, to start the lists of the {\it exposed} and the {\it active} posts and the  {\it prompted} and the {\it active} users.  
 Then at each time step:
\begin{itemize}
\item The system is driven by adding  $p(t)$ new users (Note the correspondence of one simulation step with one $t_{bin}=5$ min of real time); Their arousal and valence are given as uniform random values from  $a_i\in [0,1]$, $v_i\in [-1,+1]$, then updated with the actual mean-field terms. By the first appearance each user is given a probability $g\in P(g)$ to start a new post.  The new users are then  moved to the {\it active user list}; 

\item The emotional states 
  for all  present users are relaxed with  the rate $\gamma$,  according to the second row in the Eqs.\ (\ref{eq-arousal}) and (\ref{eq-valence});

\item The  network area  ${\cal{C}}(t,t-1)$ of the {\it active posts} is identified as post on which an activity occurred in two preceding time steps; then the lists of {\it active users} is updated from the users linked to these posts, as follows: 
  
\begin{itemize}
\item Users linked to the active posts are considered as {\it exposed} to the posted  material and decide when they will act on it, i.e.,  they are given new delay-time from the distribution $P(\Delta t)$; All users whose current delay time $\Delta t < 1 t_{bin}$ are {\it prompted} for update the emotional states  according to the first rows in Eqs.\ (\ref{eq-arousal}) and (\ref{eq-valence}), with their actual network fields computed from the Eqs.\ (\ref{eq-ha-fields}-\ref{eq-hvmf}).  An updated user is moved to the {\it active user list} with the probability  $a_0a_i(t)$ proportional to  its current arousal, else it  gets a new delay time $\Delta t \in P(\Delta t)$;
\end{itemize}
\item Every active user:
\begin{itemize} 
\item adds a new post  with the probability $g$ or otherwise comment to one of the {\it exposed posts},  which are not older than $T_0$ steps; Users are linked to posts preferentially with the probability 
$p_{p}(t)=\frac{0.5(1+v_p^{\cal{C}}(t)v_i(t))+N_p^c(t)}{\sum_p[0.5(1+v_p^{\cal{C}}(t)v_i(t))+N_p^c(t)]}$, depending on the number of comments on it  $N_p^c(t)$ and the valence similarity; 
\item and with probability $\mu$ comments a post which is older than $T_0$ steps. The post is selected preferentially according to the negativity of the  charge of all comments on it, with (properly normalized) probabilities $p_{j,old}(t)\sim {0.5+|Q_j(t)|}$, if the charge is negative, else $p_{j,old}(t)\sim 0.5$; Lifetimes of the posts are systematically  monitored (already expired posts are not considered); 

\item Current values of the valence and arousal of the  user are transferred to the posted comment or the new post; User is given a new delay-time  $\Delta t \in P(\Delta t)$; New posts are given life-time $t_P\in P(t_P)$.
 \end{itemize}

\item Delay-time $\Delta t$ for all other users is decreased by one. Time-step closes with updating the lists of the exposed and the active posts, and the lists of the exposed and the prompted users.
\end{itemize}

\subsubsection{Control parameters: Definitions and inference from the empirical data\label{sec-parameters}}
According to the above dynamic rules of the model, one can identify the parameters which control the dynamics at different levels. In particular, we use 
the following  parameters, distributions or time-series which characterize, respectively: 
\begin{itemize}
\item the local maps: $c_1=d_1=1$, $c_2=2.0$, $d_2=0.5$, $\gamma=0.05$;
\item  the properties of posts and users: $t_P\in P(t_P)$, $\Delta t \in P(\Delta t)$, $T_0=2 days$, $\mu (T_0) = 0.05$, $g\in P(g)$;
\item the driving: $\{p(t)\}$, $q=0.4$, $a_0=0.5$.
\end{itemize}

As stated above,  the {\it time-series} of the number of new users per time bin, $\{p(t)\}$, is shown in Fig.\ \ref{fig-ddDiggs_userts}.  
Several  other parameters and distributions can be also inferred from the high-resolution data of Blogs and Diggs. Specifically, in  Fig.\ \ref{fig-parameters} we show  the {\it distributions} of the  lifetime of posts, $P(t_P)$,  time delay of users actions, $P(\Delta t)$, and  the fraction of new posts per user, $P(g)$, as well as the {\it functional dependence} of the probability $\mu (T_0)$ that a user looks for a post older than the exposed posts window $T_0$. These quantities are inferred from the ddDiggs dataset.  The value $T_0=576$ time bins (corresponding to 2 days of real time) can be approximately estimated from the posts activity pattern, cf.\ Fig.\ \ref{fig-ddDiggs_patterns}b. 
The numerical values of the remaining parameters can not be extracted from this kind of empirical data. Hence they are considered as free parameters, that can be varied within theoretical limits. The values quoted above are used for the simulations in this work.

Here we describe a general methodology how such parameters are determined from the  empirical data of high resolution.
The delay-time distribution $P(\Delta t)$ in Fig.\ \ref{fig-parameters}c, is directly related with the user activity pattern, cf.\ Fig.\ \ref{fig-ddDiggs_patterns}a: for a given user (fixed index along y-axis) the delay time $\Delta t$ is defined as the distance between two subsequent points along the time axis. The distribution is then averaged over all users in the dataset. Similarly, the distribution of the life-time of posts $P(t_P)$ in Fig.\.\ \ref{fig-parameters}d,   is related to the pattern of posts activity as the distance between the first and the last point on the time axis for a given post, cf.\ Fig.\ \ref{fig-ddDiggs_patterns}b. The parameter $T_0$ is roughly estimated as the width of the time window, during which new posts were 'exposed' (dense points area in Fig.\ \ref{fig-ddDiggs_patterns}b). When $T_0$ is fixed, then the probability  that a user finds a post which is older than $T_0$ can be extracted from the data as the fraction of points beyond the dense area in the posts activity pattern until the post expires, cf.\ Fig.\ \ref{fig-ddDiggs_patterns}b. Then we have 
$\mu(T_0)=\frac{1}{N_P}\sum_{p=1}^{N_P}\left(\frac{1}{t_p}{\sum_{t_{kp}>t_{0p}+T_{0}}^{t_{p}}1}\right)$, where $N_P$ is the number of posts, $t_p$ is the expire time of the post $p$, while $t_{kp}$ and $t_{0p}$ indicate the moments of the activity at the post $p$ and its creation time, respectively.
Averaged over all posts in the dataset, gives the parameter $\mu (T_0)$, plotted in Fig.\ \ref{fig-parameters}b against $T_0$.
In the case of user properties, looking at the activity list of a given user, we can determine the fraction $g$ of new posts that the user posted out of all posts on which the user were active in the entire dataset. The values  appear to vary over time and users, the distribution $P(g)$ averaged over time and all users  in the dataset is shown in Fig.\ \ref{fig-parameters}a.  
Strictly speaking,  the values of the control parameters will depend on the empirical dataset considered. 
Specifically, the parameters as the life-time of posts, $t_P$, and users inclination to posting new posts, $g$,  or to looking towards old material $\mu(T_0)$, strongly depend on the dataset. Note also that  they might have hidden inter-dependences in view of the nonlinear process underlying the original dataset.  For instance, if on a certain Blogsite users are more inclined towards posting new material, which would yield increased probabilities of large $g$, then the life-time of posts may decline, resulting in a steeper distribution. Therefore, it is important to derive these parameters from the same dataset in order to ensure their mutual consistency.
 Note, however, that certain  universal features apply, in particular, in the power-law dependences of the delay-time \cite{crane2010,vazquez2006}  distributions $P(\Delta t)$, and the circadian cycles \cite{malmgren2009} in the time series $\{p(t)\}$.  Although our model works for a wide range of parameter values, here we keep the parameters extracted from the empirical data of the popular discussion-driven Diggs in order to enable a comparison of the results to  largest possible extent. In contrast, the relaxation rate of the arousal and the valence and the  parameters $d_1,d_2,c_1,c_2$ of the maps in Eqs.\ (\ref{eq-arousal}-\ref{eq-valence}) can not be extracted from this type of empirical data. The values shown above are chosen such that the fixed points of the maps do not fall to corner areas for typical values of the environmental fields occurring in our simulations, cf. Figs.\ \ref{fig-maps}a,b. 

\section{Results}

\subsection{Time-series of the emotional comments\label{sec-results1}}

As mentioned above, we drive the system by adding $p(t)$ new users at each time step and letting them to boost the activity of the system, according to the model rules introduced  in sec\ \ref{sec-rules}.
We sample different quantities in analogy to those that we can define and compute from the empirical data, see  section\ \ref{sec-data} and  Refs.\ \cite{mitrovic2010b,mitrovic2011}. Compared with the empirical data, the advantage of the agent-based model is that we can keep track of the fluctuations in the valence and the arousal of each agent (``user'') at all time steps.

Typically, the arousal and the valence of an agent, who is linked through the posts to other agents,  experiences stochastic inputs from the active environment, as shown in Fig.\ \ref{fig-aivi_ABMtworuns}. Between such events the emotion arousal and  valence decay with the rate $\gamma$ towards zero values.  It should be stressed that at each agent (user-node) different patterns of the activity are expected. They  depend not only on the current network structure surrounding the agent, but also on the fact that at given time the activity might be transferred to another part of the network, i.e., due to the aging of posts and the  preferences of other agents towards particular types of  posts. Two illustrative examples shown in Fig.\ \ref{fig-aivi_ABMtworuns} are from the same simulation run, but for two agents who are located at different areas of the network. 

The actions of individual agents contribute to the overall activity that can be monitored at each post and at the whole (evolving) network, as well as at the network parts, for instance the topological communities, that can be identified when the network is large enough. In the simulations we monitor the fluctuations of the number of active posts, $N_{ap}(t)$, the number of different agents that are active at these posts, $N_{au}(t)$, and the number of comments that these agents posted at each time step, $N_c(t)$. Furthermore, we distinguish between the comments that carry positive (negative) valence, $N_{\pm}(t)$, and the overall charge of these emotional comments, which is defined as the difference between the number of comments with positive and negative valence,  $Q(t)\equiv N_+(t)-N_-(t)$. The temporal fluctuations of these quantities are shown in Fig.\ \ref{fig-Psts_x2}a,b, where only the initial part of the time-series are shown, corresponding to four weeks of real time. Notice, the circadian cycles of the driving signal are reflected to the time-series of the number of active agents and the number of  their comments.

The power spectra $S(\nu)$ of these time series are  shown in the upper panels in Figs.\ \ref{fig-Psts_x2}c,d. A characteristic peak corresponding to the daily cycles of the time-series is visible. In addition,   long-range correlations  with $S(\nu)\sim 1/\nu^{\phi}$ occur in most of these time series (except for the charge fluctuations!)  for  the range of frequencies, indicated by the slopes of the straight lines in both Figures. 
The simulated time-series can be compared with the ones observed in the empirical data, for instance Fig.\ \ref{fig-ddDiggs_userts} and Figure\ S1 and with similar data analyzed in Refs.\ \cite{mitrovic2010b,mitrovic2011}. The fractality of these time series, leading to the power spectrum of the type $1/\nu^{\phi}$, as well as the dominance of the negative charge suggest that our model captures the basic features of the blogging dynamics. 
 Specifically, in response to the same driving signal, which has the power spectrum with the exponent $\phi=1.5$, 
the simulated blogging process builds the long-range correlations yielding  the time-series with  smaller exponents $\phi=1.33$ and $\phi=1$, in the number of active agents and the number of comments, respectively, and increased range of correlations, qualitatively similar to the popular Diggs. Further comparison between the simulated and the empirical data can be considered at the level of the emergent network topology, studied below in section\ \ref{sec-networks}. These time series have comprised the agent's activity at the whole (evolving) network.
In sec.\ \ref{sec-discussion} we analyze the time series which are recorded at the level of each emerging agent-community separately.

\subsection{Emergent network and  communities of the emotional agents\label{sec-networks}}
The networks that emerge through the activity of our emotional agents on posts can be studied in full analogy with the bipartite networks mapped from the empirical  data of the same structure, which are discussed in section\ \ref{sec-data} and in Refs.\ \cite{mitrovic2010a,mitrovic2010b,mitrovic2010c}. (For other interesting examples and review of complex monopartite networks, see \cite{boccaletti2006} and references therein). 
In our simulations the network evolves due to the addition of nodes of both partitions, as well as the evolution of links. The lists of user---posts connections are updated at every time step.  With the prevailing negative comments, as demonstrated above in the time-series, cf. Fig.\ \ref{fig-Psts_x2}, arrivals of the negative comments at posts generate an environment that, in view of the linking rules (large negative charge preference), strongly affects  the network evolution  (see also discussion in sec.\ \ref{sec-discussion}).  In this way some posts with  a large  number of negative comments and thus large topological strength may appear. 
A part of the emergent network obtained in our simulations  is shown in Fig.\ \ref{fig-bipnetABM}, where three such hubs---popular posts are visible together with the users linked to them. 

 Some  topology measures of the emergent bipartite network---the degree distributions and the assortativity measures,  are shown in Fig.\ \ref{fig-bipnetABMtopology}a,b. For comparison, the corresponding topology measures of the network obtained from the empirical dataset of discussion-driven popular Diggs are also computed and shown in supporting information, Fig.\ S3. 
As expected, the degree distributions for each of the partitions---agent(user)-nodes and post-nodes appear to be different (some other examples of bipartite networks representing the empirical data of various  techno-social interactions have been analyzed in \cite{mitrovic2010c}). Specifically, the broad distributions are dominated by different type of cut-offs. They can be approximated by the following mathematical expressions, motivated by fitting the corresponding empirical data, cf. Fig.\ S3: 
\begin{equation}
P(q_u) =C_uq_u^{-\tau}e^{-q_u/X_{u0}} ;~~~~~~~~~~  P(q_p)=C_p\left[1-(1-q)\left(\frac{q_p}{X_{p0}}\right)\right]^{1/1-q} ;
\label{eq-distributions} 
\end{equation}
for the agent(user)-node  and the post-node distributions, respectively. The agent-degree distribution $P(q_U)$ exhibits a short power-law region and a large exponential cut-off, whereas the distribution related to the post-nodes $P(q_P)$ has a dominant cut-off at smaller degree followed by a power-law tail, compatible with the $q$-exponential form with $q>1$. The fitted values of the parameters are shown in the Figure legends in Fig.\ \ref{fig-bipnetABMtopology}a. In principle, they depend on the simulation parameters of the agent-based model. However, the expressions appear to be stable with respect to the simulation time (size of the network). The results are shown for two simulation runs with 16384 and 25000 time steps,  resulting in the networks with  $N_P=13504 + N_U=64852$, and  $ N_P=22757 + N_U=107933$ nodes, respectively.
 The same mathematical expressions in eq.\ (\ref{eq-distributions}) apply (with different parameters)  to the corresponding distributions derived from the empirical data, cf. Fig.\ S3a. Hence, in these measures the topology of the bipartite network emerging in the emotional blogging of our model shares qualitative similarity with the one from the popular posts in real data. 
To certain extent, similar conclusions apply in the case of the mixing patterns, shown in Fig.\ \ref{fig-bipnetABMtopology}b and in Fig.\ S3b. Namely, the posts making the  network neighborhood of an agent-node (user-node in the empirical data) exhibit no assortativity measure, which is indicated by the line of zero slope  before a cut-off at large user-node degree. In the case of post-nodes, the empirical data indicate slight disassortativity (decrease) just before the cut-off, Fig.\ S3b, a feature that seems not  be properly captured by our model in  the present parameter range.
Systematic study of the network topologies, emerging  when  the  parameters of  blogging dynamics within the agent-based model are varied in a wide range away from their empirical values, is left for a separate work  \cite{MMBT_abmnetworks}.
Here we focus on the mesoscopic structure of the emergent network obtained for the current set of parameters, which are listed above. In particular, we are interested in communities of the emotional agents, that may potentially occur on the respective  monopartite projection of the network.

Communities,  as topological subgraphs with stronger connections between the nodes inside the community compared to the rest of the network,  can be accurately identified using different methods
\cite{rosvall2008,evans2009,mitrovic2009,fortunato2010}. The bipartite networks representing the dynamics on Blogs and Diggs  exhibit abundance of different communities  \cite{mitrovic2010a,mitrovic2010b, mitrovic2011}. A systematic analysis of such  bipartite networks, representing various online interactions, can be found in \cite{mitrovic2010c}. 
The monopartite user- or post-projections of these networks appear to be highly clustered and weighted networks \cite{mitrovic2010a,mitrovic2010b, mitrovic2011,mitrovic2010c}, which limits applications of classical methods of community analysis \cite{fortunato2010}. Therefore, we use the methods based on the eigenvalues spectral analysis of the network \cite{donetti2004,mitrovic2009} and the maximum likelihood method adapted for multi-graphs \cite{mitrovic2008}. 
The networks emerging through the actions of the emotional agents in our simulations have similar features. 
 
We perform the spectral analysis of the normalized
Laplacian operator   \cite{dorogovtsev2003,mitrovic2009} which is related to the weighted user-projection network, whose matrix elements ${C_{ij}^W}$ represent the common number of posts per pair of users, {\it including the  multiplicity of user--post connections}, which is  indicated by the superscript. It is constructed from the symmetric matrix of {\it commons} as
\begin{equation}
 L_{ij}=\delta_{ij}-\frac{C_{ij}^W}{\sqrt{(l_{i}l_{j})}} \ ,
\label{eq-Laplacian}
\end{equation}
where $l_{i}$ is the \textit{strength} of node $i$, defined as the sum of wights of its links. 
As discussed in detail in Refs.\ \cite{dorogovtsev2003,mitrovic2009}, the spectrum of the Laplacian\ (\ref{eq-Laplacian}) is limited in the range $\lambda_i\in[0,2]$. When the communities exists, the  lowest non-zero eigenvalues of the Laplacian\ (\ref{eq-Laplacian}) appear separated from the rest of the spectrum and the corresponding eigenvectors are localized on the network subgraphs (communities). The localization of the eigenvectors is visualized as a characteristic branched structure of the scatter-plot in the space of these eigenvectors. This property of the eigenvectors is then utilized to identify the nodes of the network that belong to each community (detailed discussion and various  examples studied by this methods can be found in \cite{mitrovic2009,grujic2009,mitrovic2010a,mitrovic2011}). 
It should be stressed that the network grown through the emotional actions of of the agents in our model has specific properties which may be reflected in the community structure. Namely, the bipartite network is already weighted, which shifts  the distribution of weights in the monopartite projection away from  pure topology of commons \cite{mitrovic2010c}. In addition, the network evolves in such way that the center of the activity is shifting to ever new groups of (exposed) posts. 

Here we analyze the structure of the network after 4032 time steps (two weeks) of the evolution. The network projected onto user (agents) partition contains 
$N_{U}=4572$ users, only users with the degree larger than $5$ are considered as relevant for the community formation. The eigenvalue spectrum of the Laplacian operator Eq.\ \ref{eq-Laplacian} with the $C_{ij}^W$ matrix  related with this  user-projected weighted network, is then computed. The results for the eigenvalues shown in the ranking order are given  in Fig.\ \ref{fig-spectra}a. The scatter-plot of three eigenvectors belonging to the three lowest eigenvalues is shown in Fig.\ \ref{fig-spectra}b.
The spectrum, as well as the scatter-plot  in Fig.\ \ref{fig-spectra}b, indicate that five agent-communities can be differentiated. These are denoted by $G_k$, with $k=1,2,\cdots 5$ corresponding to top-to-bottom branches in Fig.\ \ref{fig-spectra}b. In the following we first identify the nodes representing the agents in each of these  communities. Then we analyze how the communities actually evolved on that network and discuss the fluctuations of the emotional states of each agent in the communities through the evolution time. 

 The  communities on user-projected networks are our main concern, in view of the collective behavior of the emotional agents. Formally, the same methodology can be applied to the post-projected network as well as the weighted bipartite network directly. Studies of the empirical data \cite{mitrovic2010a,mitrovic2010b}  suggest that  the communities of posts often appear in relation to their subjects, age, and sometimes authorship. The features  is prominent  in the case of Blogs of ``normal'' popularity. Whereas, in the communities on popular posts the subjects are often mixed, leaving potentially different driving force for user's intensive activity on that posts.  
In our model the post have no defined subjects, and the agents are driven by the emotion content alone. Nevertheless,  the results of the spectral analysis of the post-projected network reveals that several communities of the posts can be identified. The network projection is described in supporting information and the scatter-plot of the respective eigenvectors is shown in  Fig.\ S4. Four communities can be differentiated. Looking at the node's identity in these branches, we find that the ``age'' of the posts prevails as a grouping principle. Other attributes of posts that we have in the model, as ``authorship'', ``popularity'' and ``charge'', seem to be mixed in all present communities.

\section{Discussion\label{sec-discussion}}
\subsection{Patterns of agents behavior in time and space of the emotion variables}

Having identified the agents in each of the communities, we can track of their group activity and the emotion fluctuations over time from our simulation data.  
The time-series of the number of comments of all agents in a given community $G_k$ are shown in Fig.\ \ref{fig-communities}a and the emotional charge of the valence of these comments---in the Fig.\ \ref{fig-communities}b.
Note that a fraction of comments with the valence values close to zero in the range  between $(-0.01,+0.01)$ are considered as neutral, and do not contribute to the charge.
The profile of the time series indicates that all communities  
started to grow at early stages of the network evolution.  
However, two of them, $G_1$ and $G_5$, ceased to grow and reduced the activity relatively quickly after their appearance.  Looking at the fluctuations of   charge of the emotional comments in these two communities, we find that it is well balanced, fluctuating around zero at early times, and eventually leveling up to zero.  
Whereas, in the other two medium-size communities, $G_2,G_4$,  the activity is slowly decreasing, while the largest central community, $G_3$, shows constantly large activity.  
Comparing the activity (number of comments) with the fluctuations in the charge of the emotional comments, we can see that in these three communities
 the excess negative charge settles  after some time, breaking the initial balance in the charge fluctuations. 
In this way our model reveals the correlations between the prolonged activity and the size of a community  (i.e., number of different agents), on one side, with the occurrence of the negative charge of the related  comments, on the other, a feature also observed in the empirical data on Blogs and Diggs \cite{mitrovic2010b,mitrovic2011}. 

 In view of the preference towards the posts with negative charge, a comment regarding the breaking of the charge balance and its consequences to the network topology is in order. Note that, according to the model rules, cf. sec.\ \ref{sec-model}, the probability of a post to receive first negative/positive comment depends on the valence of the agent who is active on that post and its similarity with the average valence of the currently active posts. 
Contrary to the naive preference towards node's degree, which is known to lead to a scale-free degree distribution in growing monopartite networks (for theoretical derivation and the conditions when power-law  distributions occur, see Ref.\ \cite{SD_prl2001}), in our model agents preference is driven by another quality of a post node---its emotional content. Hence, in this process no scale-free distributions of the posts degree is expected, as also shown above. More importantly, the negative charge appears to have limited fluctuations. The time-series of the (negative) charge and of the number of comments remains  stationary over large periods of time, before the activity ceases, as shown in Figs.\ \ref{fig-communities} for the communities, and in Fig.\ \ref{fig-Psts_x2}b for the entire network.

Another interesting feature of these communities can be observed by visualizing the patterns of their activity in  the {\it phase space of the  emotion variables}. In order to match the emotion measures accepted in the psychology literature, i.e., according to the 2-dimensional Russell's model \cite{russell1980,scherer2005}, we  use the circumplex map suggested in Ref.\ \cite{ahn2010}.  
The values of the arousal and the valence variables  are thus mapped onto a surface enclosed by a circle, where, 
according to Refs.\ \cite{scherer2005,russell1980}, the emotions commonly known, as for instance, ``afraid'', ``astonished'', ``bored'', ``depressed'', etc., can be  represented by different points (or segments) of the surface. 
In particular, the values of the arousal and valence are mapped as follows \cite{ahn2010}:
\begin{equation}
 a^{'}=\frac{ a^{1}}{\sqrt{1+z^{2}}} \ , v^{'}=\frac{v}{\sqrt{1+z^{2}}} \ ,
\label{eq-circumplex}
\end{equation}
where $z\equiv min{(\vert a^1\vert ,\vert v\vert )}$ and  $a^1 \in [-1,+1]$ is obtained from our arousal variable by first mapping $a^1\equiv  2a-1 $. 

Computing the transformed values of the arousal and the valence for each agent in a given community at all time steps when an action of that agent is recorded in our simulations, we obtain the color-plots shown in  Figs.\ \ref{fig-circumplex4groupsABM-RunII}. 
Specifically, the color map indicates  how often a particular state on the circumplex was occupied in  four of the above communities, normalized with the all actions in that community. As the Figure\ \ref{fig-circumplex4groupsABM-RunII} shows, the communities leave different patterns in the space of emotions.
For instance, the community $G_1$,  that have balanced charge fluctuations, appears to cover a larger variety of the emotional states, leading to the pattern on the top left figure. 
Whereas, when a large community is formed, it  may induce large negative fields which keep the agents in the negative valence area of the circumplex map.  The situation corresponding to  the community $G_3$ is shown in bottom left plot in Fig.\ \ref{fig-circumplex4groupsABM-RunII}. Majority of the comments is this case are centered in the area of the arousal and the valence where the negative emotional states known as  ``worried'', ``apathetic'', and  ``suspicious'', ``impatient'', ``annoyed'' etc, are found  on the circumplex map (see  Refs.\ \cite{scherer2005,ahn2010} for coordinates of some other well known emotional states covered by these patterns).
Plots  on the right-hand side of Fig.\ \ref{fig-circumplex4groupsABM-RunII} correspond to the communities $G_2$ and $G_4$, in which charge fluctuations are moderately negative, as discussed above.

From the Figures \ref{fig-circumplex4groupsABM-RunII} we can also observe that  the well defined lower bound for the agent's arousal emerges in  a self-organized manner inside each community, although no sharp threshold exists in the model rules. Moreover, the arousal drives the valence when the agents are active. This is clearly displayed in the case of communities with a balanced charge, as our community $G_1$, Fig.\ \ref{fig-circumplex4groupsABM-RunII} top left. Similar pattern of the arousal--valence was  found in the laboratory experiments \cite{bradley2001}, where the  values of the arousal and valence are inferred from skin conduction, heart beat and facial expression measurements on users reading a selection of  posted texts. 

 The characteristic patterns  in Figs.\ \ref{fig-circumplex4groupsABM-RunII} emanating from the emotional blogging of our agents suggest the processes with anomalous diffusion, in which certain parts of the phase space are more often visited than the others. They reflect  the self-organized dynamics of the agent's emotion variables and the network topology. Formally, these patterns are between two extreme situations: synchronous behavior, focusing at a lower-dimensional areas, and random diffusion, spreading evenly over the entire  space.
 The most visited areas of the phase space are in the vicinity of the attractors of the nonlinear maps. The positions of these attractors for each agent map move,  depending on the actual values of the fields acting on it. The fields themselves fluctuate over time for each map, being tunned by the the agent's emotion variables and the local topology of the network (community) where the agent is situated.

\subsection{Conclusions\label{sec-conclusions}}
In this work we have adapted  the idea of the emotional agents with two-dimensional emotion states following Ref.\ \cite{schweitzer2010}, and implemented them onto a networked environment with a bipartite network of agents (users) and posts, to model the dynamics on Blogs and similar Web portals,  where the emotions are communicated indirectly via posts.  In addition to the emotions, measured by the dynamical variables of the arousal and the valence of each agent, appearance of our agents on the network follows circadian cycles and and the action-delay from a distribution, which is, like some other parameters of the dynamics, inferred from the empirical data.

Novel and the most important features of our agent-based model are: 
\begin{itemize}
\item the agent's  emotion is spreading via indirect contacts through their actions on the  posts on a bipartite network, and 
\item the network itself evolves  through the  addition of the agents and their actions on the posts.
 \item We also present a systematic methodology to extract the relevant parameters of the model from the empirical datasets of high temporal resolution.
\end{itemize}
Apart from the inference of the control parameters from a given empirical dataset, the structure of the model allows experimenting  on the dynamics and potential extensions at three different levels: 
(a) changing the properties of each individual agent; (b) modifying the linking rules, i.e., posts exposure, agent's preferences, role of the emotions, etc.; (c) introducing different and/or additional driving of the system, and varying the balance between the influences of the local and the global events. 

In this work the focus was on the following aspects of the dynamics:
\begin{itemize}
\item the emergence of the collective states due to agent's  emotional communications;
\item the patterns of their emotional behaviors  in time and in phase space of emotion variables;
\item the role of underlying contacts and the structure of emergent network. 
\end{itemize}
For this purpose, in the simulations we consider no external input to drive the system except for the agents arrivals according to the time signal $\{p(t)\}$, inferred from  the empirical data of the discussion-driven popular Diggs. We have demonstrated that   the communities of agents emerge through the emotional commenting of posts,  and can be identified as topological subgraphs on the weighted user-projected network. Most of the activity of the (user) agent occurs inside the community where the agent belongs. Moreover, the growth of the community is self-amplified and prolonged  with the excess negative charge of the related comments. 
Another quantitative measure of the collective behavior is found in  building the correlations in the streams of events---$1/\nu ^{\phi}$-type of the power spectrum for the time series is found for the emotional comments,  in the response to the driving signal, which has weaker correlations.  Moreover, this also applies for the time-series of comments with positive/negative emotion valence.
These features are in {\it qualitative agreement} with those found  in the empirical data from the discussion-driven Diggs and Blogs, analysed by the same mathematical approaches \cite{mitrovic2010c,mitrovic2011},  cf. also Fig.\ \ref{fig-ddDiggs_userts}. Hence we can conclude that the dynamic rules of our agent-based model  reveal the key mechanisms behind the collective emotional behaviors, which are observed on the popular posts in real Blogs and Diggs.

To examine the role of the driving signal in building the collective response, we perform comparative simulations in which the features of the $p(t)$ signal are completely ignored. Instead, we drive the system by adding a {\it constant number $p=6$ of agents per time step}. The simulated time-series are shown in supporting information, Fig.\ S5a and its power-spectra in Fig.\ S5b. Apart from higher average activity and the absence of daily cycles, the fractality of the time-series is preserved. The power-spectrum with the exponent $\phi=1.25$ is observed for the number of comments, although in a smaller range. The negative charge sets-in after certain time period and fluctuates in a stationary manner. These results suggest that the occurrence of long-range temporal correlations is inherent to the stochastic process of our model, which might be  enhanced, but not imported, by the profile of the driving signal. It is also interesting to point that the network topology grown in this process exhibits the degree distributions of the same type as Eq.\ (\ref{eq-distributions}) and three communities of the emotional agents, as shown in Fig.\ S5c and d.

Compared with the empirical data, where the emotion is extracted only from the text on posts, in the agent-based model we can follow the fluctuations of the emotion of each agent over time. In this respect our model can interpolate between the psychology experiments at individual users, on one side, and the global emotional states, that can be recognized at larger scale \cite{dodds2009}, on the other. 
We have shown that the activities within each  community of the agents leave a characteristic pattern in the space of emotions. Specifically, ``normal'' blogging leads to  balanced emotions in a segment of the circumplex map, driven by large arousal (i.e.,   between the lines marking  the limits of ``high power/control'' and ``obstructive'' emotions in Ref.\  \cite{scherer2005}). This is in agreement with the laboratory measurements of the emotional states obtained on individual users \cite{bradley2001,thelwall2010}. 
However, their collective behaviors may be entirely different.
Our simulations suggests how the large communities can get caught into excessive negative emotions (critique), which prolongs the activity and the number of agents involved. The model provides the underlying mechanisms and the parameters through which such collective behavior  may be influenced.

 In a broader view of the science of affective computing, our work aims to quantitative accounting of the {\it dynamics of emotions}. With the rules and parameters motivated by the discussions on popular Diggs, our agent-based model describes the dynamic environment with indirect communications among users, where the critique prevails over positive emotions.  Keeping track of the relations between simulated events in terms of an evolving bipartite network is another aspect of this work, which makes the basis for the quantitative analysis of agents collective behaviors. 
Furthermore, the network with its local structure is essential for the discussion type dynamics that we study. By taking away the network of posts, the agents would be literally disconnected and their dynamics greatly simplified, consisting of a single appearance-and-relaxation event per agent. The role of the network structure can be altered by increasing the mean-field term (i.e., its strength $q$ and the selection of posts which contribute to it), or by adding an external noise to Eqs.\ (\ref{eq-arousal}-\ref{eq-valence}) which acts directly to each agent, as in Ref.\ \cite{schweitzer2010}.
Qualitative agreement of the simulation results with the ones of the empirical dataset indicates how the emotional communications prevail on popular posts, leading to the bursting events and the emergence of communities. Whereas, the quantitative differences, in particular in the user heterogeneity shown in Fig.\ S3 compared to the the degree distribution of our emotional agents in Fig.\ \ref{fig-bipnetABMtopology}, suggest to what extent features other than emotion may drive users behavior. 
Potentially different outcomes can be predicted by the model when parameters are changed or extracted from another dataset. On the other hand,  when the posts of  ``normal'' popularity are considered, we expect that subjects of the posts  may play a role, affecting the way that certain emotions are communicated. Such situations can be studied by appropriate modifications of the dynamic rules  within our model and inference of the related parameters.

In conclusion, compared to extracting user's emotional behavior from texts of Blogs and Diggs by quantitative analysis of the empirical datasets, the agent-based modeling has the advantage in that the emotional processes are studied at each agent (representing user) directly. Moreover, variety of social experiments can be devised and performed on the agents, thus avoiding the ethical issues related with the real users.
Despite its mathematical  complexity, our agent-based model  obviously represents  simplified reality both in respect to the dimensionality of emotions and the agent properties. For instance, in the science of emotion it is known that  social aspects of the emotions such as ``guilt'', ``pride'', ``shame'' are different  compared to ``anger'', ``depression'', ``joy'', and others. The psychological theory also suggests that personality profile determines how certain emotion is expressed.  
The theoretical modeling along the lines of our agent-based model will benefit from current developments in the science of affective computing, which aims for quantitative measures behind the psychology theory and for extracting different dimensions of emotions \cite{calvo-review2010}. The agent properties can also include more realistic personal differences and influence of the real-world processes on users decisions and rhythm of their stepping into the virtual world. 
In the present model such real-world processes are {\it implicitly}  taken into account  through the parameters---the driving signal $\{p(t)\}$ and the delay-time distribution $P(\Delta t)$, which are involved in generating the respective empirical data of Diggs. This is a step beyond the first approximation,  which we hope is opening the way for further research towards  realistic modeling of the emotion dynamics on the Web. 
\\

\section{{\bf Figure Legends and  Supporting Information}}


{\bf Figure 1:}{\bf Patterns of activity differ for user and post nodes.} Example of temporal patterns of (a) user actions and (b) activity at posts, obtained  from the original dataset of discussion-driven Diggs (ddDiggs). Indexes are ordered by the user (post) first appearance in the dataset, while time is given in minutes.

{\bf Figure 2:}{\bf Time-series with circadian cycles and fractal features.} (bottom) Time-series of the number of new users (red-dark) and the number of all active users  (green-pale) per time bin of 5min derived from the ddDiggs dataset;  (top) Power spectra of these time series as indicated (shifted vertically for better vision). Daily and weekly cycles can be easily noticed on both plots.

{\bf Figure 3:} {\bf Two-dimensional maps of the emotion variables of an isolated user-node.} Maps for arousal (left) and valence (right) for four different values of the fields are shown. The fixed line  $X(t+1)=X(t)$ is indicated.

{\bf Figure 4:}{\bf Parameters of the model as inferred from the empirical data of ddDiggs.}
(a) Distribution of $g$---the fraction of new posts per user, relative to all posts on which that user was active, averaged over all users in the dataset. (b) Probability $\mu$ that a user looks at a post which is older than the specified time window $T_0$ time bins, averaged over all users and plotted against $T_0$. (c) Distribution of the time-delay $\Delta t$ between two consecutive user actions, averaged over all users in the dataset. (d) Distribution of the life-time of posts $t_P$, averaged over all posts in the dataset (log-binned data). In Figs. (c) and (d) the time axis is given in the number of time bins, each time bin corresponds to $t_b=5$ minutes of real time.

{\bf Figure 5:}{\bf Valence and arousal of each agent linked on the network fluctuate in time.} Two examples of the valence and the arousal are shown against time for two agents  (users) located in different areas of the bipartite network, resulting in different activity patterns: a very active  agent (left) and a sporadically active agent (right).

{\bf Figure 6:} {\bf  Time series simulated in the model of interacting emotional agents on bipartite network.} The number of all comments per time step (cyan) and the number of comments with positive (red) and negative (black) valence per time step (a), and  the double-logarithmic plot of their power spectra (c).   The number of active posts (indigo) and the number of active agents (magenta)  per time step, and the charge of all  emotional comments (blue), are shown in panel (b), and their corresponding power spectra, panel (d). Straight lines indicate slopes $\phi$=\{1, 1.33, 0\}. For clear display, the power-spectra are logaritmically binned.

{\bf Figure 7:} {\bf Emergent bipartite network of the emotional agents and the posts exhibits strong inhomogeneity.} Shown is a  part of the network structure in the vicinity of three popular posts.

{\bf Figure 8:} {\bf Some topology measures of the bipartite network emerging in the dynamics of emotional agents.}
(a) The degree distributions of the user-partition ($\bigcirc$,$+$) and the post-partition ($\square$,$\times$). Fitting lines explained in the Legend.  (b) Assortativity measures: The average degree of the posts linked to the user node of a given degree versus user degree ($\bigcirc$,$+$), and the average user degree linked to the post of a given degree, plotted against post degree ($\square$,$\times$). Empty symbols are for the simulation time 16384 steps, while the crosses indicate the respective results from runs with 25000 time steps.

{\bf Figure  9:} {\bf Spectral analysis of the emergent network reveals community structure.} For the agent-projected network (a) the eigenvalue spectrum, and (b) the scatter-plot of three eigenvectors belonging to the lowest nonzero eigenvalues, indicate  five communities  $G_k$, $k=1,2 \cdots 5$, related to five  branches in the scatter-plot, from top to bottom. Note that each point in the scatter-plot represents a unique node on the network.

{\bf Figure  10:} {\bf Active communities grow in correlation with the excess of negative charge.} Time series of the number of comments by the agents belonging to a given community $G_k$ (a), and the charge of these comments (b), computed for each of five  communities identified in Fig.\ \ref{fig-spectra}.

{\bf Figure 11:} {\bf Activity patterns of the communities projected onto 2-dimensional space of emotions.} Circumplex map of the emotional states of the agents belonging to four communities identified on the emergent network: $G_1$--top left, $G_2$--top right, $G_3$--bottom left, $G_4$--bottom right.
 Color map indicates occupancy of a given state, normalized relative to the number of comments in each community.
\\

{\bf Figure S1:} {\it Prevalence of negative comments on popular Diggs.} 
Shown are the time-series and their power-spectra of the number of comments $N_\pm(t)$ carrying  positive emotion (red) and negative emotion (black), and the time-series of ``charge'' of the emotional comments, $Q(t)=N_+(t)-N_-(t)$ (cyan). The Figure shows that comments are correlated in time, leading to $1/\nu$-type of power spectra and that the negative comments prevail.

{\bf Video S2:} {\it User interests shift daily towards new posts.}
The movie shows the evolution of activity of one community on the  weighted bipartite network of discussion driven Digg stories during $90$ days. 
In Ref.\ \cite{mitrovic2011}, we have identified such communities in the network constructed from the subset of the discussion-driven Diggs and  keeping only very active users (with more than 100 comments). The network is then analyzed by the eigenvalue spectral methods\cite{mitrovic2009}  and three communities are identified \cite{mitrovic2011}.
 For the movie, we selected nodes belonging to one of these communities, $g_{2}$, which consists of  $236$ users and posts, and then determined weighted bipartite subnetworks of it for each consecutive day. Thus, the weight of the link corresponds to the number of comments written by the user on the post  {\it on a given day}, while the color of the link indicates overall emotional contents of these comments (black--negative, white--neutral and red--positive).  The networks are then visualized using the program {\it Pajek}. Every frame corresponds to the window of one day in real time. The frames are combined using {\it Avidemux} program package. The movie shows that on each day different posts were in the focus.

{\bf Figure S3:} {\it Some topology measures of the bipartite network derived from empirical data on Diggs.} For comparison of the network topology, the dataset of the popular discussion-driven Diggs, described in section\ \ref{sec-data}, is mapped onto a weighted bipartite network, with the weights of links representing the number of comments of a user to a post. 
The degree distribution of user-nodes and post-nodes of that network are computed and  shown in Fig.\ S3(a), 
while the assortativity measures---the average degree of the post nodes linked to the user of a given degree, and vice versa, the average degree of the user nodes linked to the  post of a given degree, are shown in Fig.\ S3(b). For completeness, shown are also the respective quantities computed from whole dataset of all Diggs (including the posts with normal popularity), indicated in the Legend.

{\bf Figure S4:} {\it Community structure of post-projected network obtained from the emotional agents dynamics.} Shown is the scatter-plot of the eigenvectors belonging to three lowest eigenvalues of the post-projected weighted bipartite network, emerging in the simulations  of the emotional agents dynamics after 4000 time steps. The network of size is reduced to $N_P=1156$ posts by taking the posts with strength $\ell _P>50$, as relevant for the community formation.

{\bf Figure S5:} {\it When the bloggers never sleep.}
Simulations results obtained for the system of the emotional agents is driven by adding a constant number of users $p=6$ per time step: Time-series of the number of comments and charge (a)  and their power-spectra (b). Degree distributions of user-nodes and post-nodes, $P(q_U)$ and $P(q_P)$, are shown in panel (c).Scatter-plot of the eigenvectors related with the user-projected network ($N_U=4418$ users with strength $\ell_U>10$ considered), indicating the community structure (d).
\\

{\bf Funding} The research leading to these results has received funding from the European Community's Seventh Framework Programme FP7-ICT-2008-3 under grant agreement 
n$^o$ 231323 (CYBEREMOTIONS). B.T. thanks support from the  national program P1-0044 (Slovenia). 

{{\bf Author Contributions} Conceived and designed the experiments: BT. Performed the experiments: MM. Contributed analysis tools: MM. Analyzed the data: MM, BT. Wrote the paper: BT.


\
\begin{figure}[h]
\resizebox{38.0pc}{!}{\includegraphics{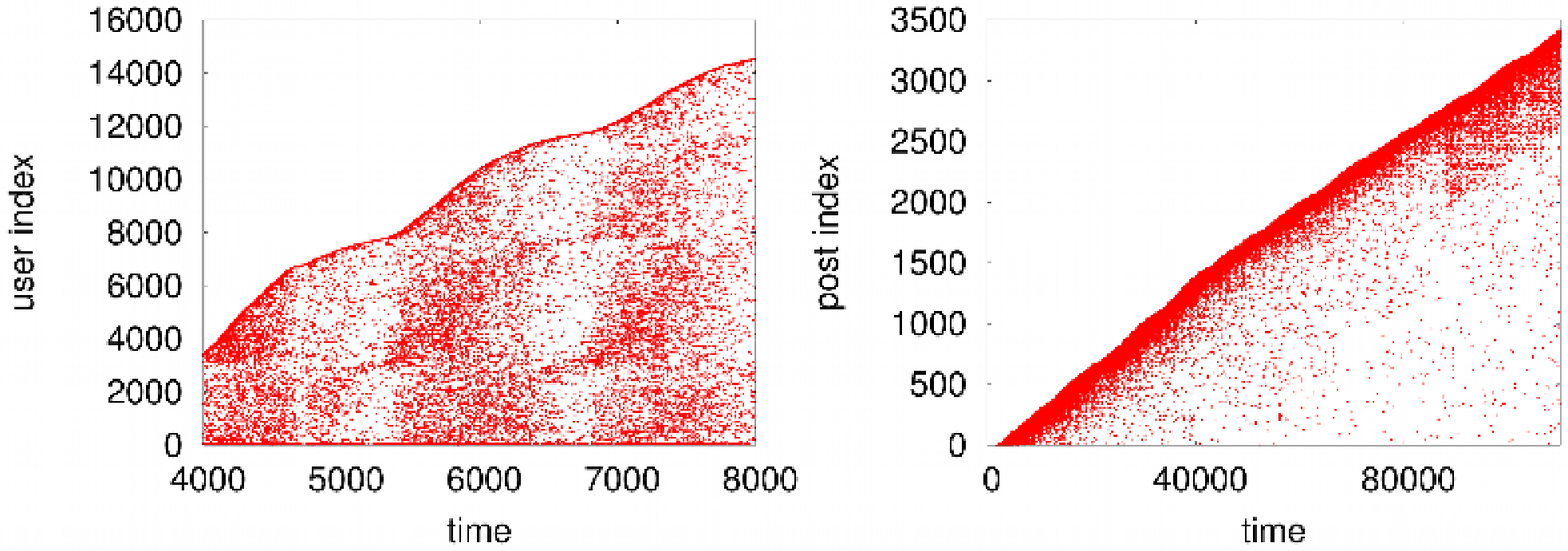}}
\caption{}
\label{fig-ddDiggs_patterns}
\end{figure}

\begin{figure}[h]
\centering
\resizebox{38.0pc}{!}{\includegraphics{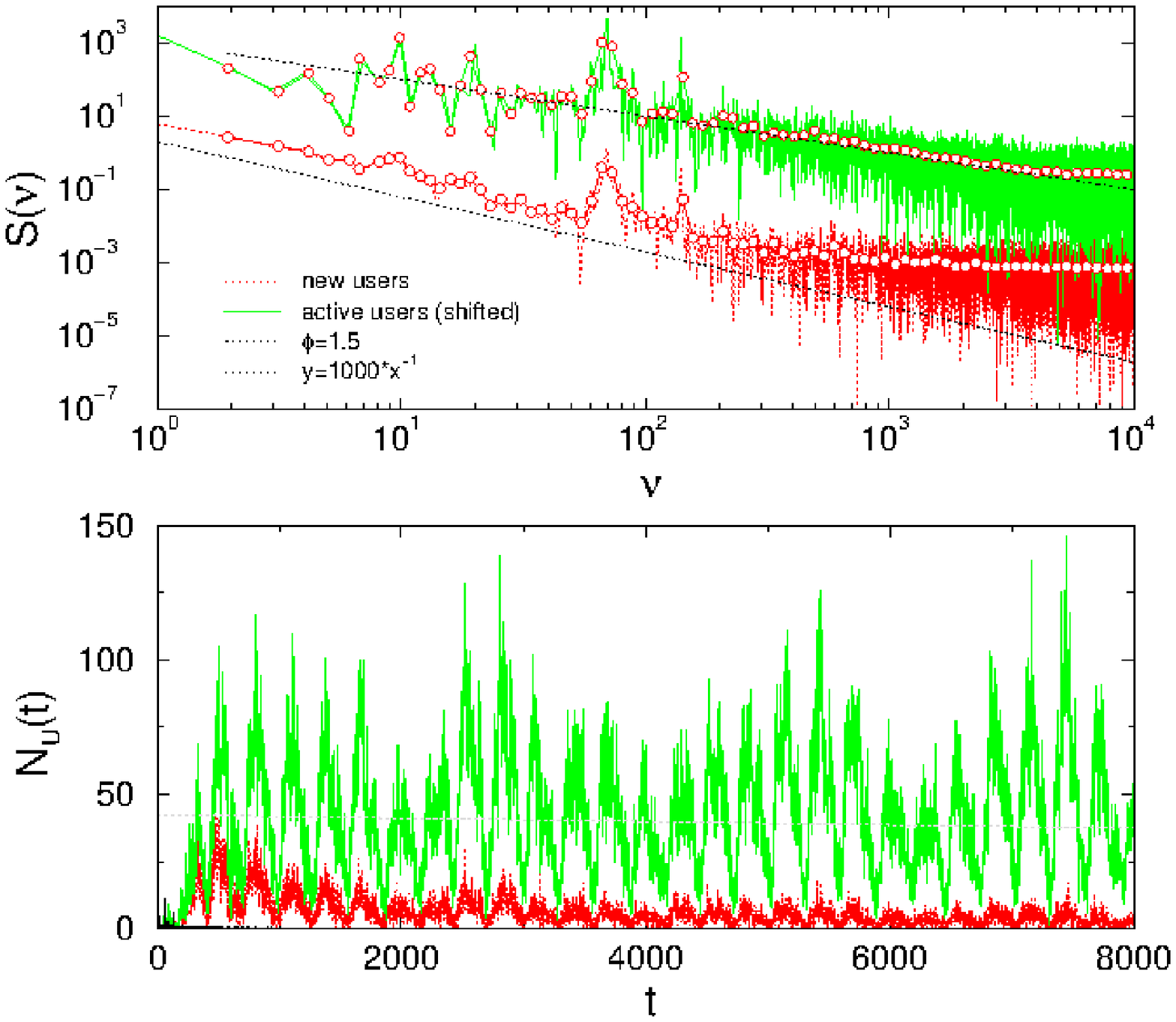}}\\
\caption{}
\label{fig-ddDiggs_userts}
\end{figure}

\begin{figure}[h]
\resizebox{38.0pc}{!}{\includegraphics{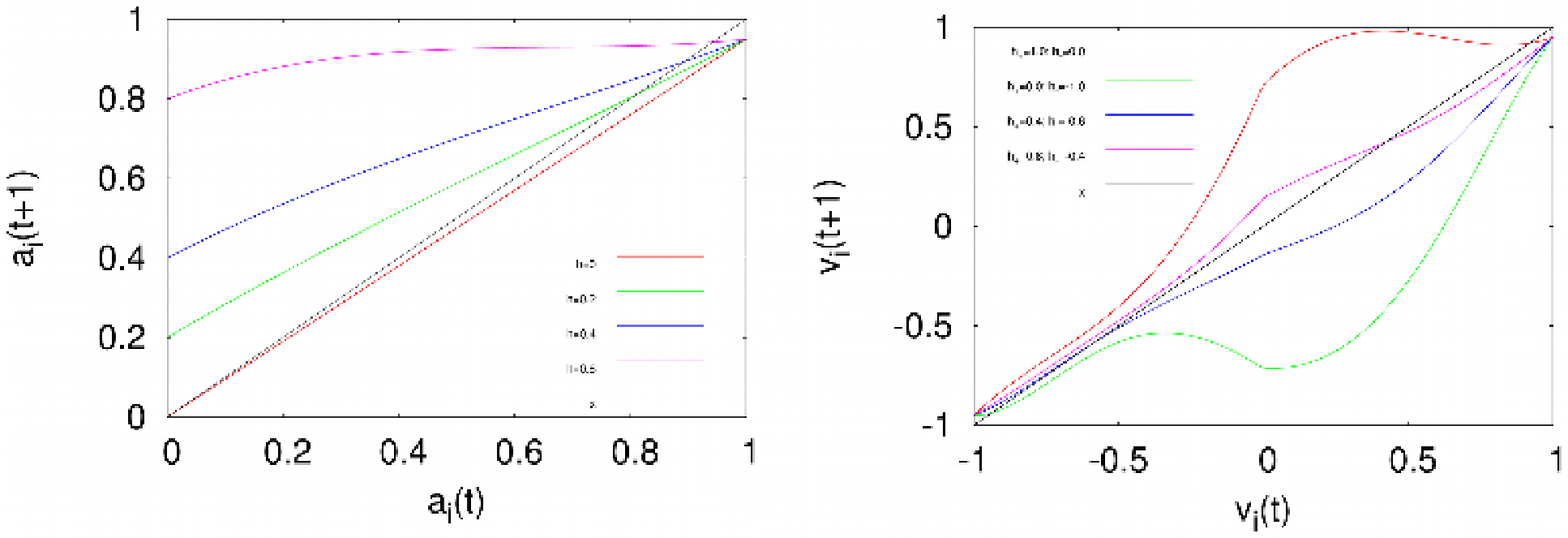}}  
\caption{}
\label{fig-maps}
\end{figure}

\begin{figure}[h]
\centering
\resizebox{38.0pc}{!}{\includegraphics{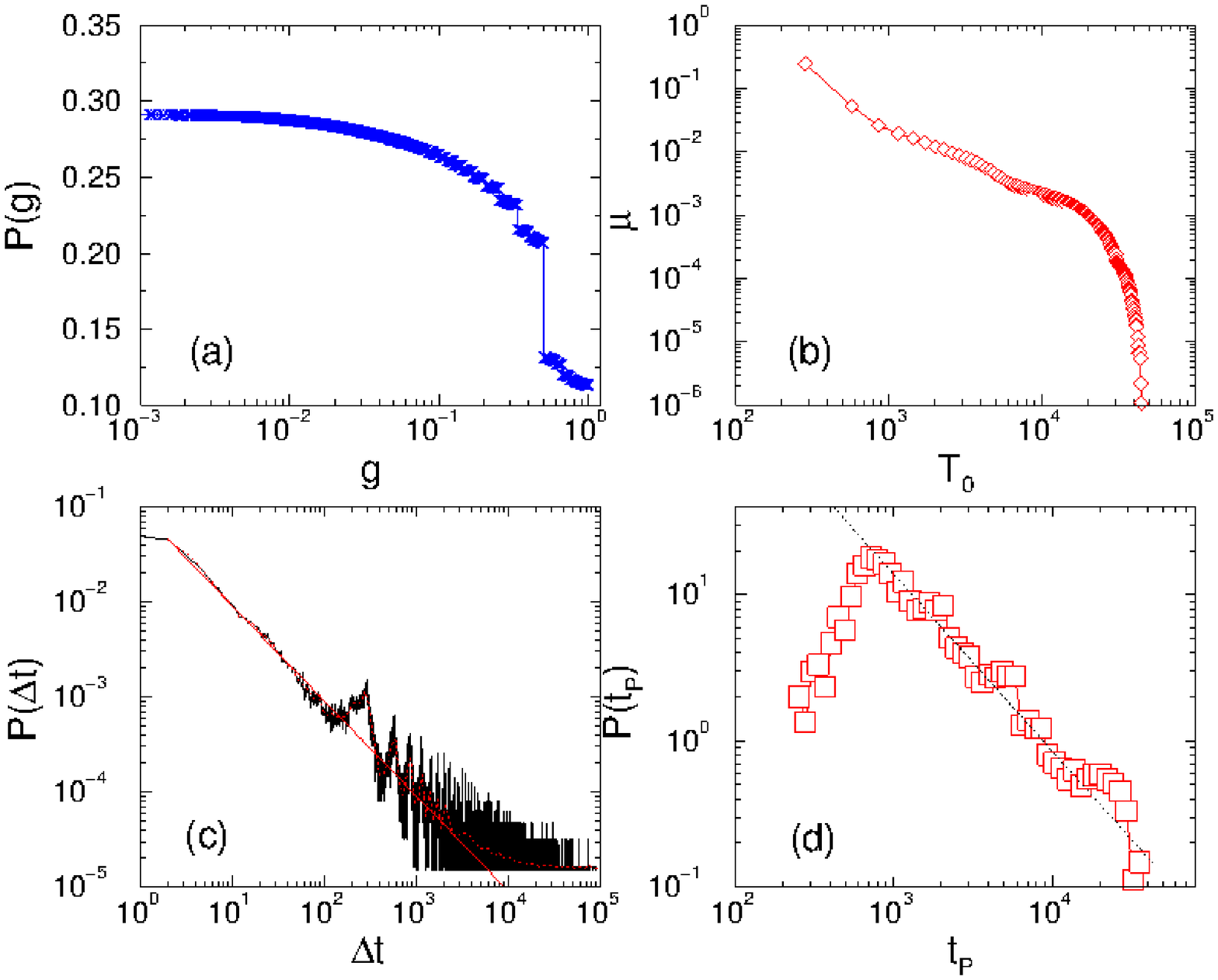}}\\
\caption{ } 
\label{fig-parameters}
\end{figure}

\begin{figure}[h]
\centering
\resizebox{38.0pc}{!}{\includegraphics{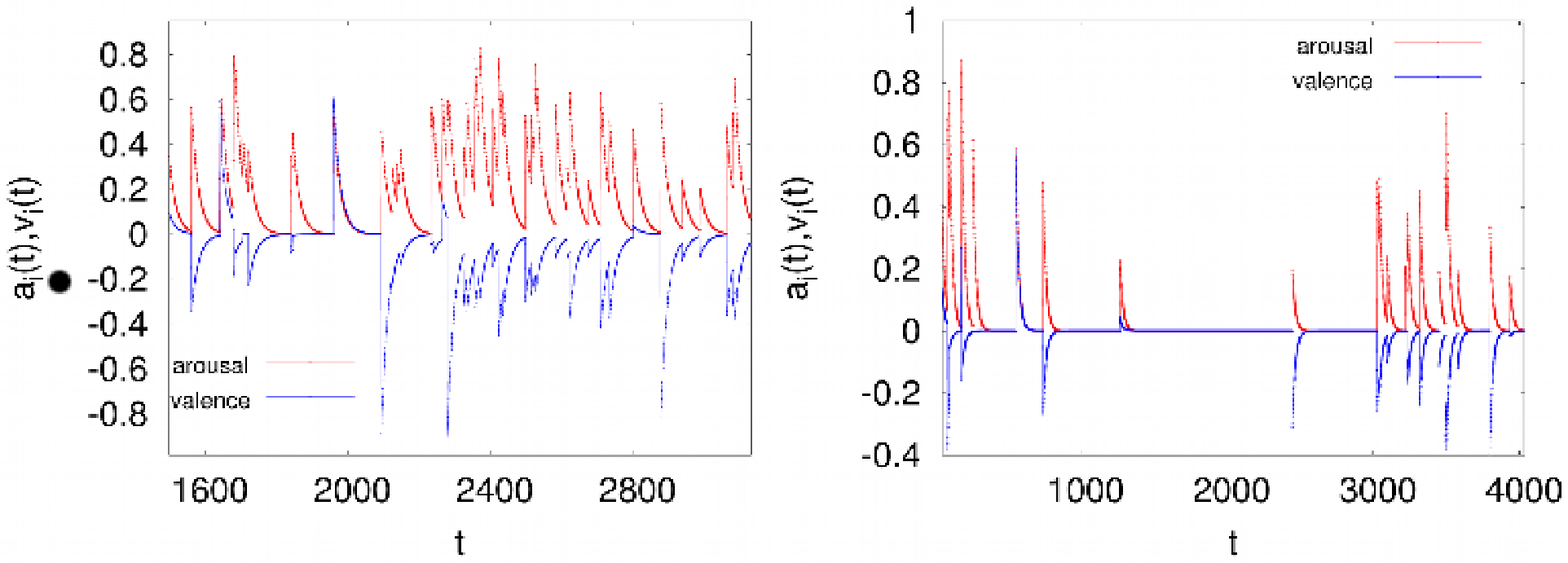}}\\  
\caption{}
\label{fig-aivi_ABMtworuns}
\end{figure}

\begin{figure}[h]
\centering
\resizebox{38.0pc}{!}{\includegraphics{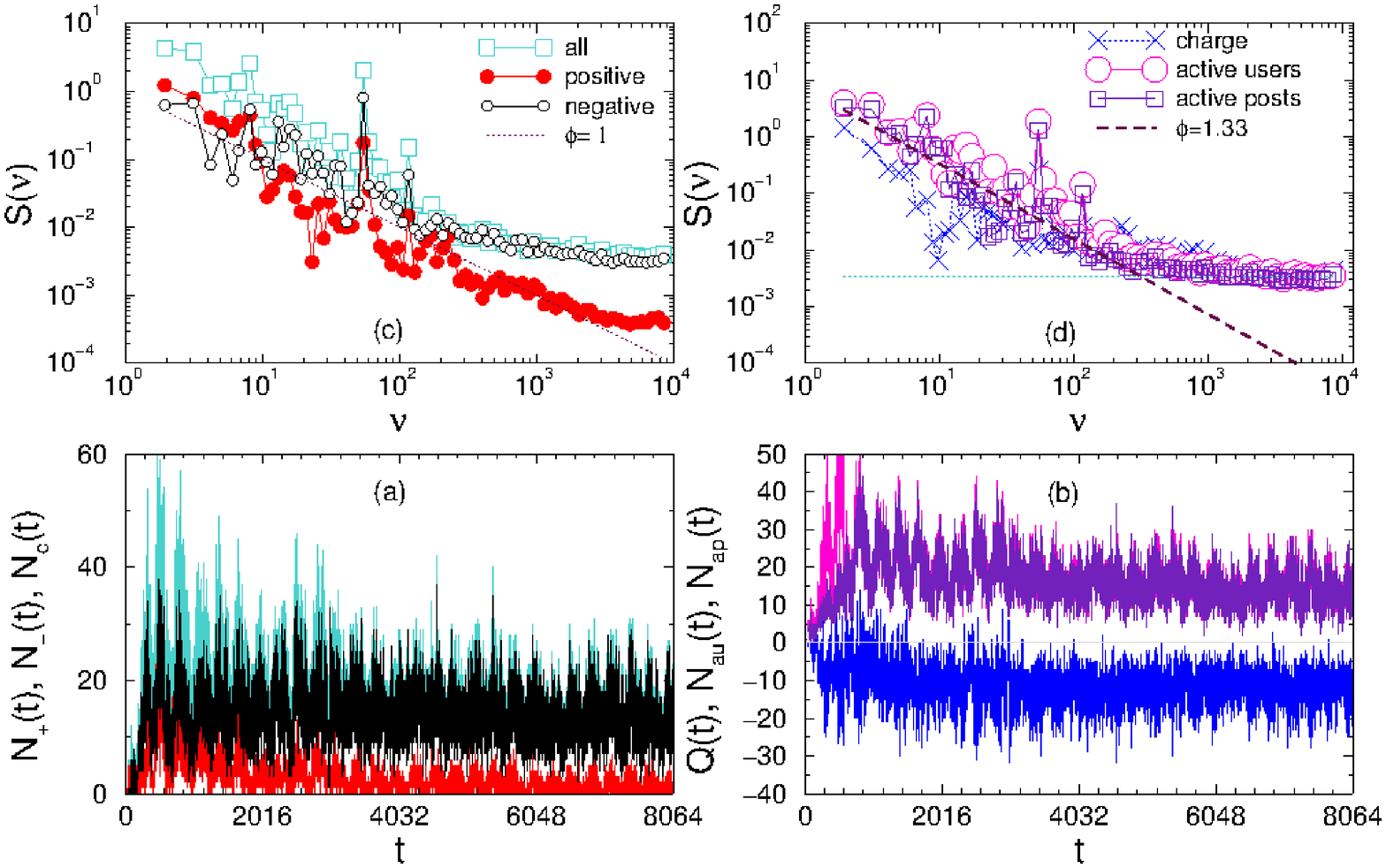}}\\  
\caption{}
\label{fig-Psts_x2}
\end{figure}

\begin{figure}[h]
\centering
\resizebox{24.8pc}{!}{\includegraphics{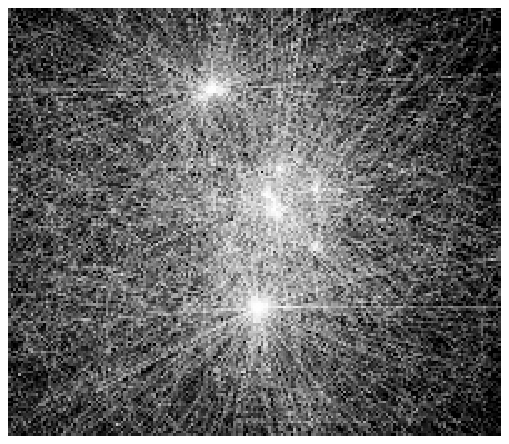}}\\  
\caption{}
\label{fig-bipnetABM}
\end{figure}

\begin{figure}[h]
\centering
\resizebox{28.8pc}{!}{\includegraphics{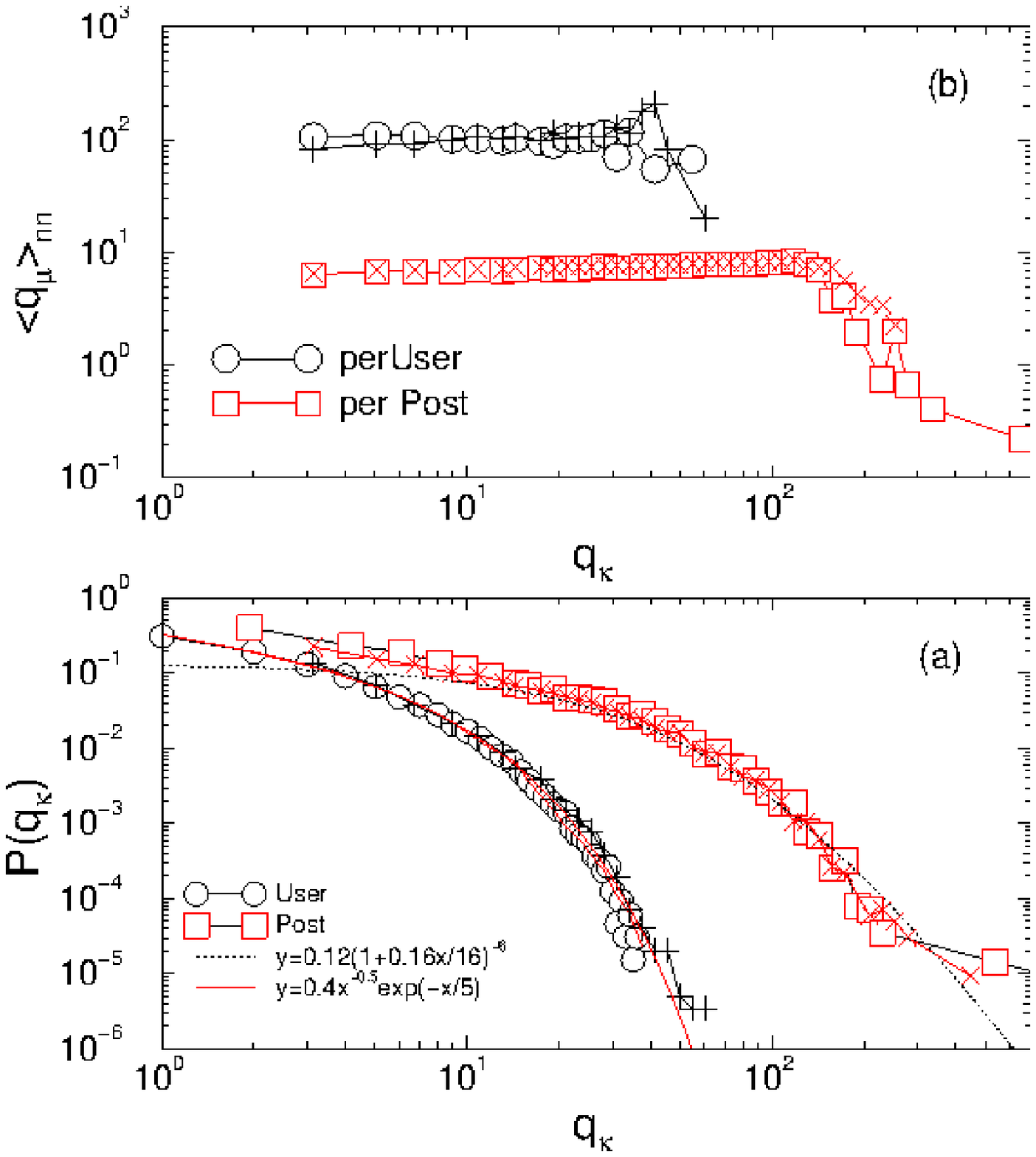}}\\
\label{fig-bipnetABMtopology}
\end{figure}

\begin{figure}[h]
\centering
\resizebox{38.0pc}{!}{\includegraphics{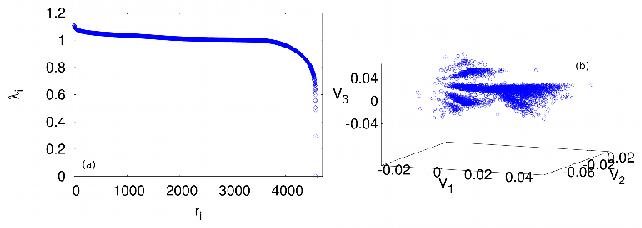}}\\
\caption{ }
\label{fig-spectra}
\end{figure}

\begin{figure}[h]
\centering
\resizebox{38.0pc}{!}{\includegraphics{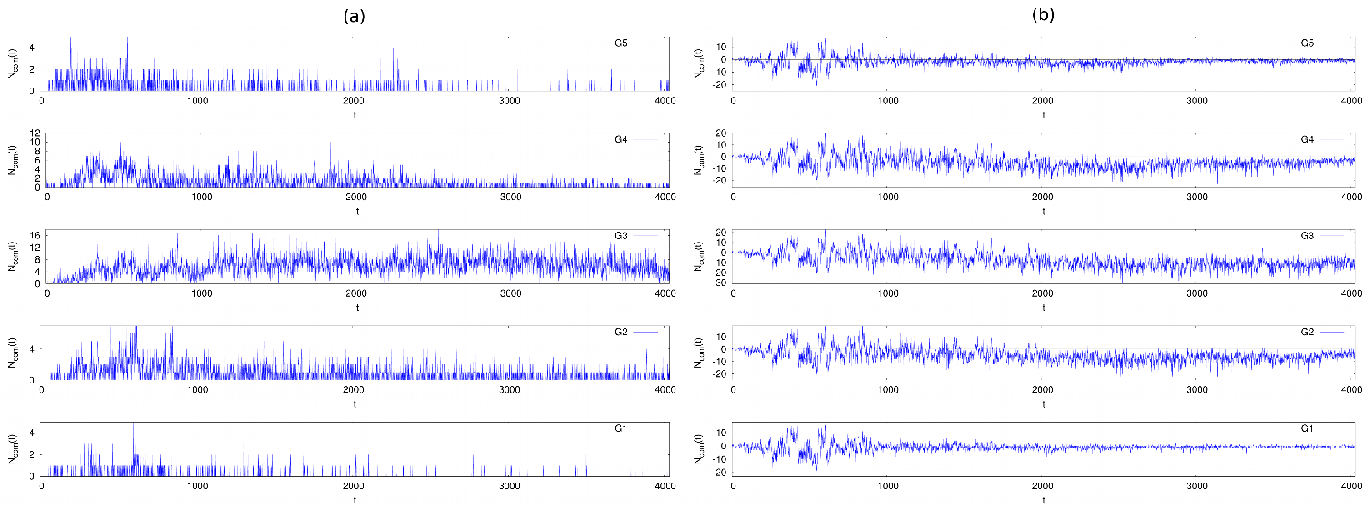}}\\  
\caption{}
\label{fig-communities}
\end{figure}

\begin{figure}[h]
\centering
\resizebox{38pc}{!}{\includegraphics{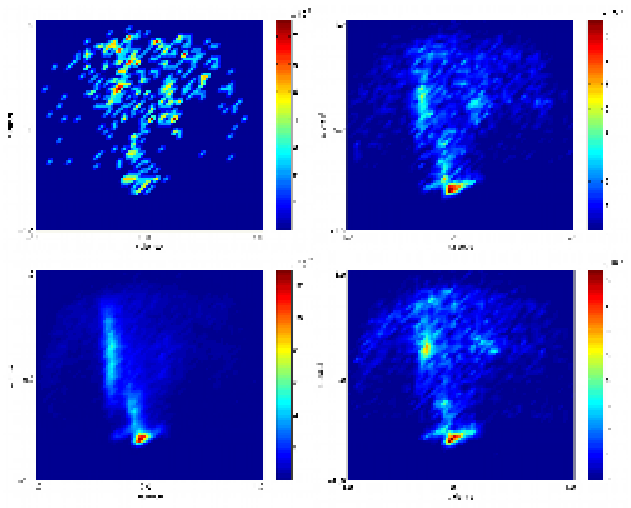}}
\caption{}
\label{fig-circumplex4groupsABM-RunII}
\end{figure}
\begin{figure}[h]
\centering
\resizebox{38pc}{!}{\includegraphics{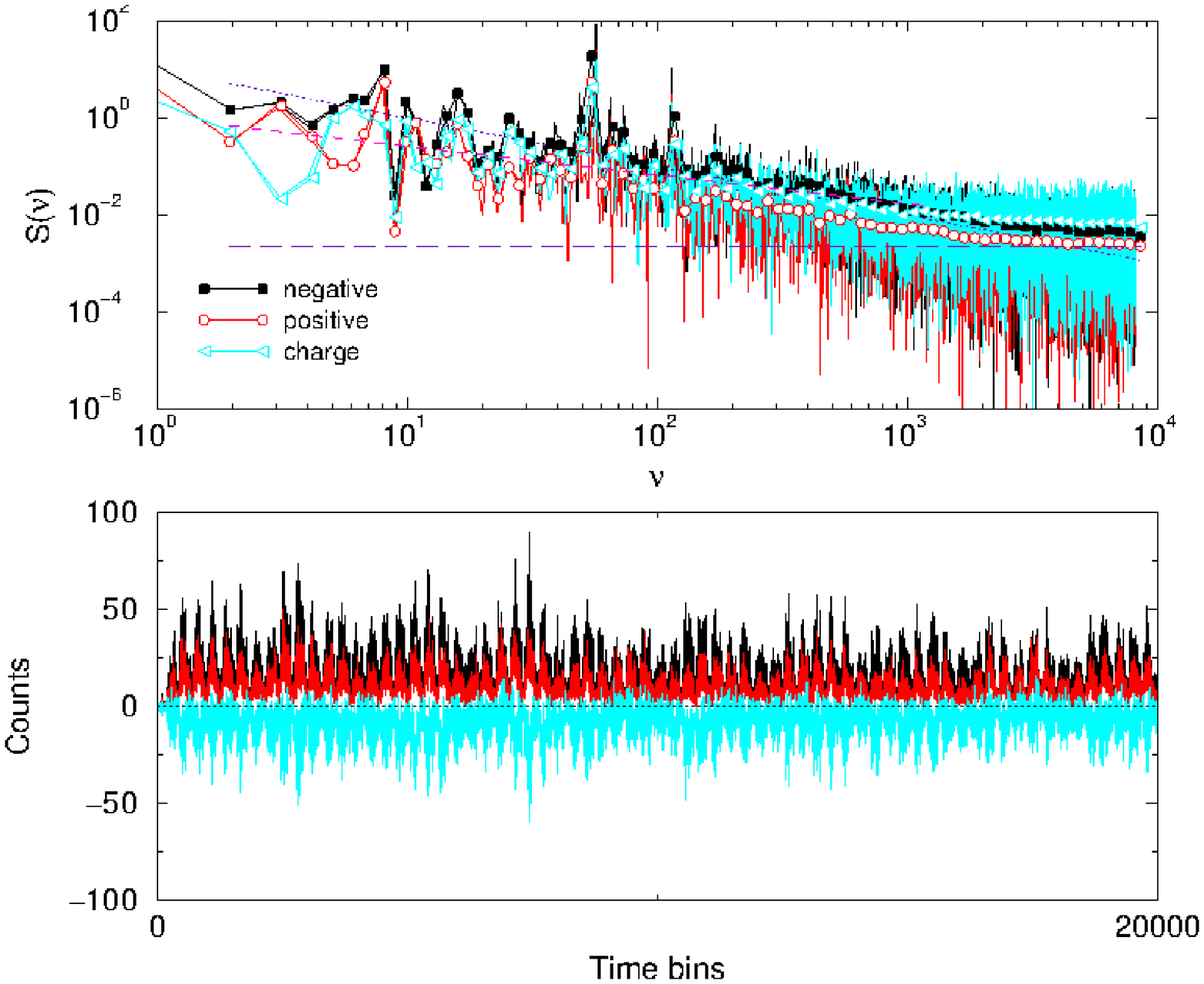}}
\caption{S1}
\label{fig-si_S1}
\end{figure}
\begin{figure}[h]
\centering
\resizebox{38pc}{!}{\includegraphics{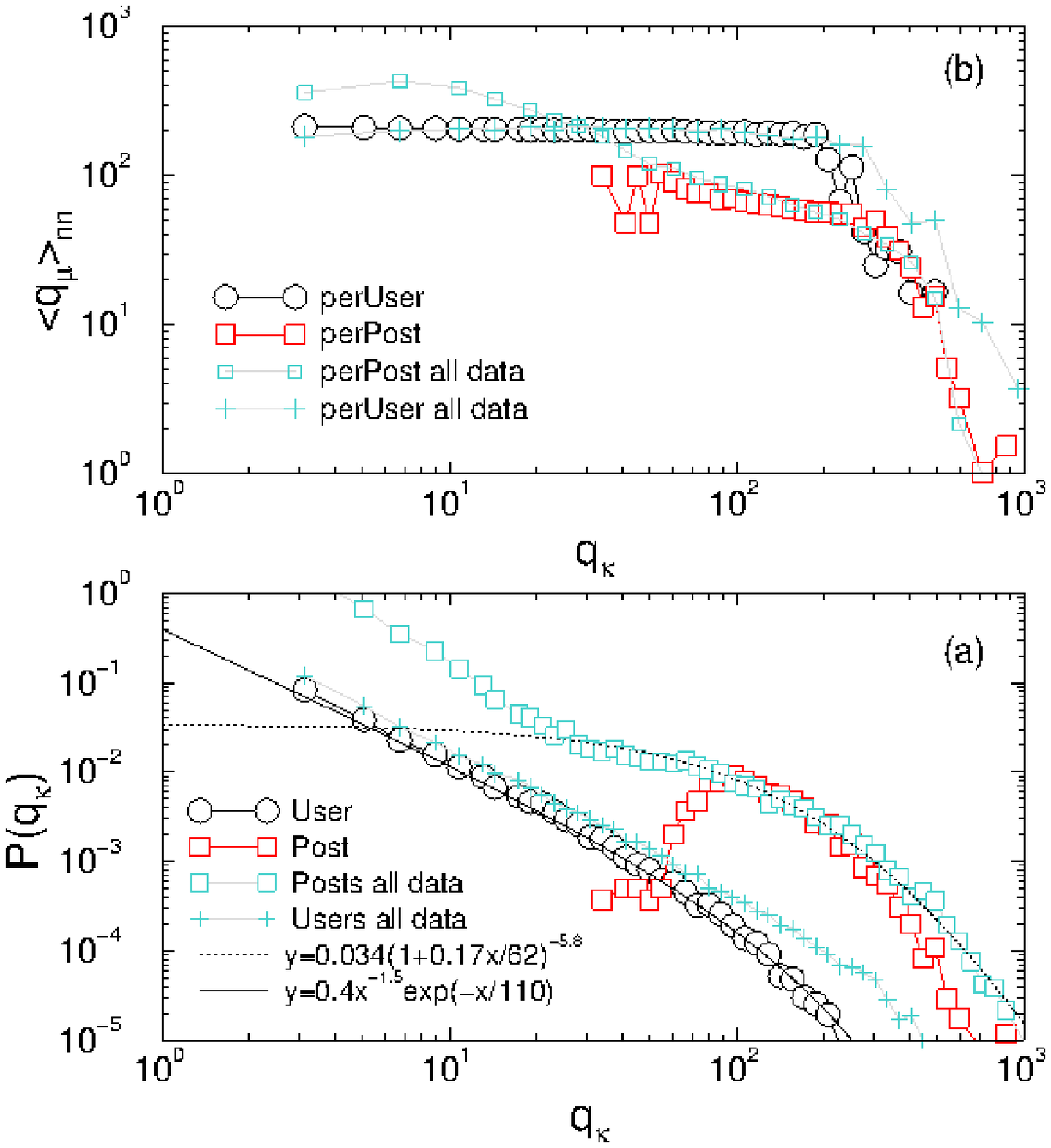}}
\caption{S3}
\label{fig-si_S3}
\end{figure}
\begin{figure}[h]
\centering
\resizebox{38pc}{!}{\includegraphics{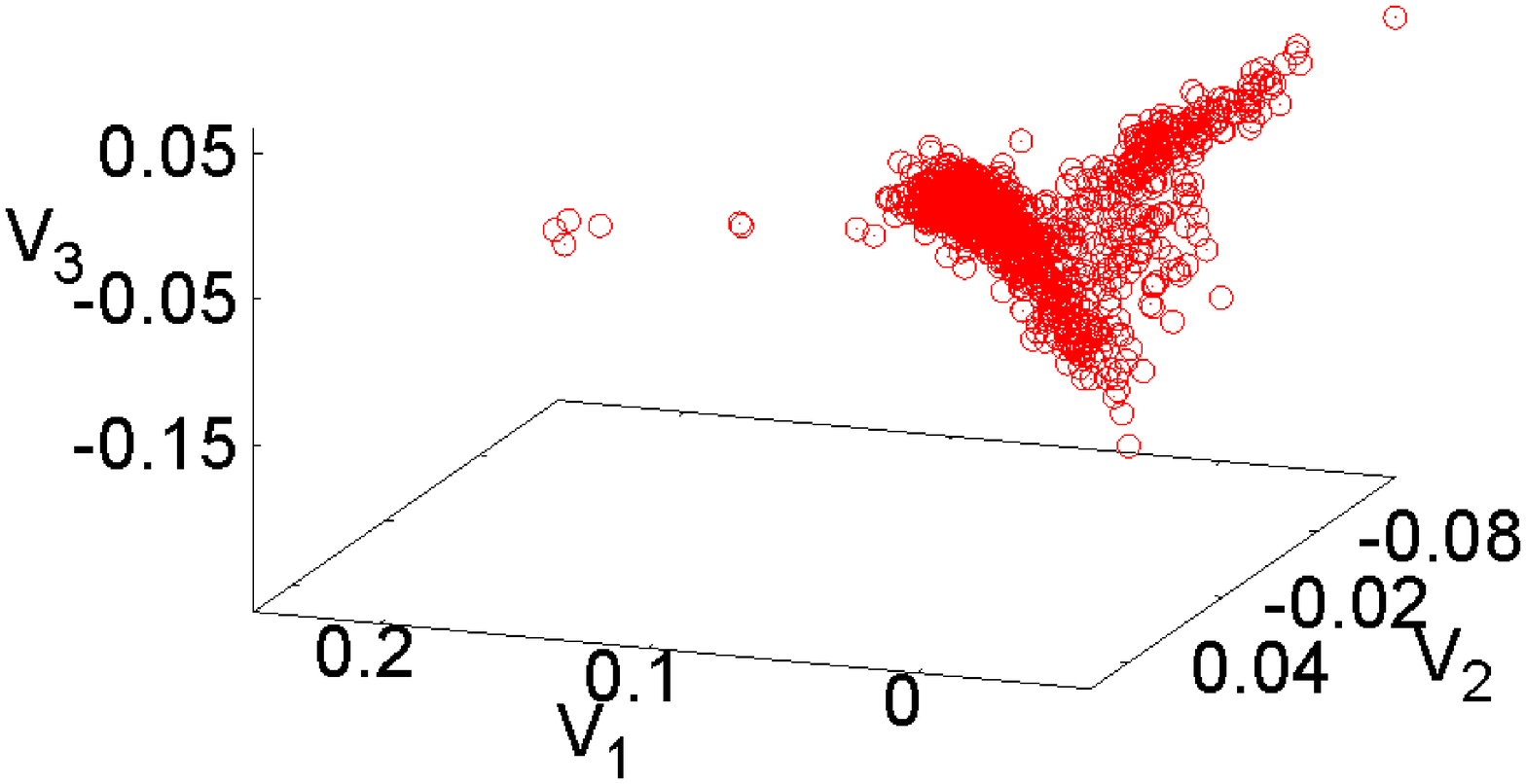}}
\caption{S4}
\label{fig-si_S4}
\end{figure}

\begin{figure}[h]
\centering
\resizebox{38pc}{!}{\includegraphics{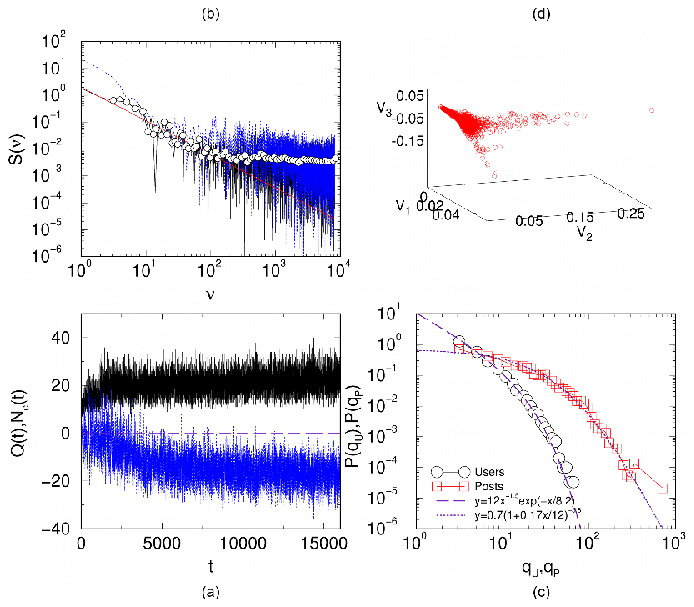}}
\caption{S5}
\label{fig-si_S5}
\end{figure}

\end{document}